\pdfoutput=1
\documentclass[10pt]{iopart}

\usepackage[colorlinks=true,linkcolor=blue]{hyperref}
\usepackage{amsfonts,latexsym,eucal,amssymb} 
\usepackage{graphicx}
\usepackage{psfrag}
\usepackage{color}
\usepackage{amsbsy}
\usepackage{amssymb}       
\usepackage{multicol}
\usepackage{wrapfig}
\usepackage{blindtext}
\usepackage{lineno}
\usepackage{comment}
\usepackage[square,numbers,sort&compress]{natbib} 
\usepackage{braket}
\usepackage{fullpage}
\usepackage[export]{adjustbox}
\usepackage{tikz} 

\def\dd{\mbox{d}}

\def\q{{\bf q}}

\def\x{{\bf x}}
\def\y{{\bf y}}

\def\t{{\bf t}}

\def\p{{\bf p}}

\def\Q{{\bf Q}}
\def\S{{\bf S}}

\def\HS{\hspace{0.5mm}}

\begin{document}

\title[PI]{A Path Integral Approach to Age Dependent Branching Processes}

\author{Chris D. Greenman}

\address{School of Computing Sciences, University of East Anglia, Norwich, UK, NR4 7TJ}

\begin{abstract}
Age dependent population dynamics are frequently modeled with generalizations of the classic McKendrick-von Foerster equation. These are deterministic systems, and a stochastic generalization was recently reported in \cite{Greenman2016, Chou2016}. Here we develop a fully stochastic theory for age-structured populations via quantum field theoretical Doi-Peliti techniques. This results in a path integral formulation where birth and death events correspond to cubic and quadratic interaction terms. This formalism allows us to efficiently recapitulate the results in \cite{Greenman2016, Chou2016}, exemplifying the utility of Doi-Peliti methods. Furthermore, we find that the path integral formulation for age-structured moments has an exact perturbative expansion that explicitly relates to the hereditary structure between correlated individuals. These methods are then generalized with a binary fission model of cell division.
\end{abstract}

\vspace{2pc}
\noindent{\it Keywords}: Doi-Peliti, Second Quantization, Path Integral, Birth-Death Process, Age-Structure

\maketitle

\noindent\rule{16cm}{0.4pt}

\tableofcontents

\noindent\rule{16cm}{0.4pt}



\section{Introduction}

Populations are dynamic entities that exhibit stochastic variation through time; the number of individuals in the population fluctuates. Examples range from microscopic populations of cells to evolutionary processes of large multicellular organisms \cite{Charlesworth1994}, and models of the underlying processes find a range of application. Historically these models have been deterministic, such as with Malthusian exponential growth. However, for populations of smaller volume, or for those going through a bottleneck, fluctuations can have significant effects on the dynamics and stochastic properties need to be considered.

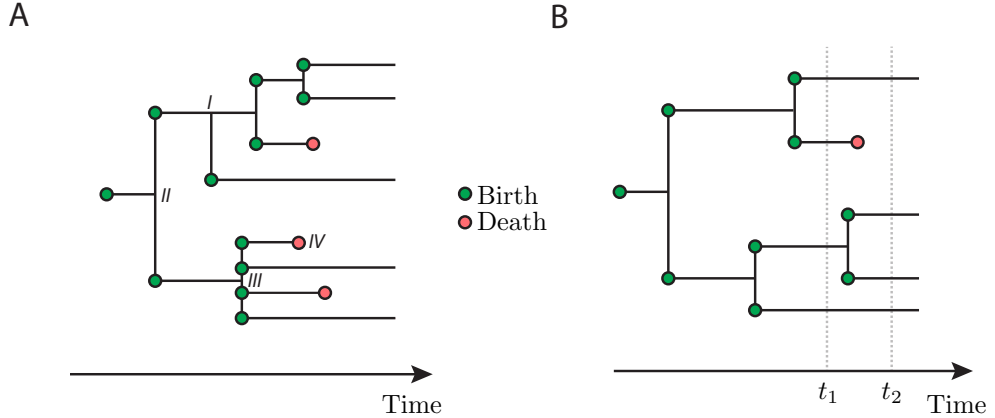
\begin{figure}[t]
\hspace{28mm}
\includegraphics[width = 13cm]{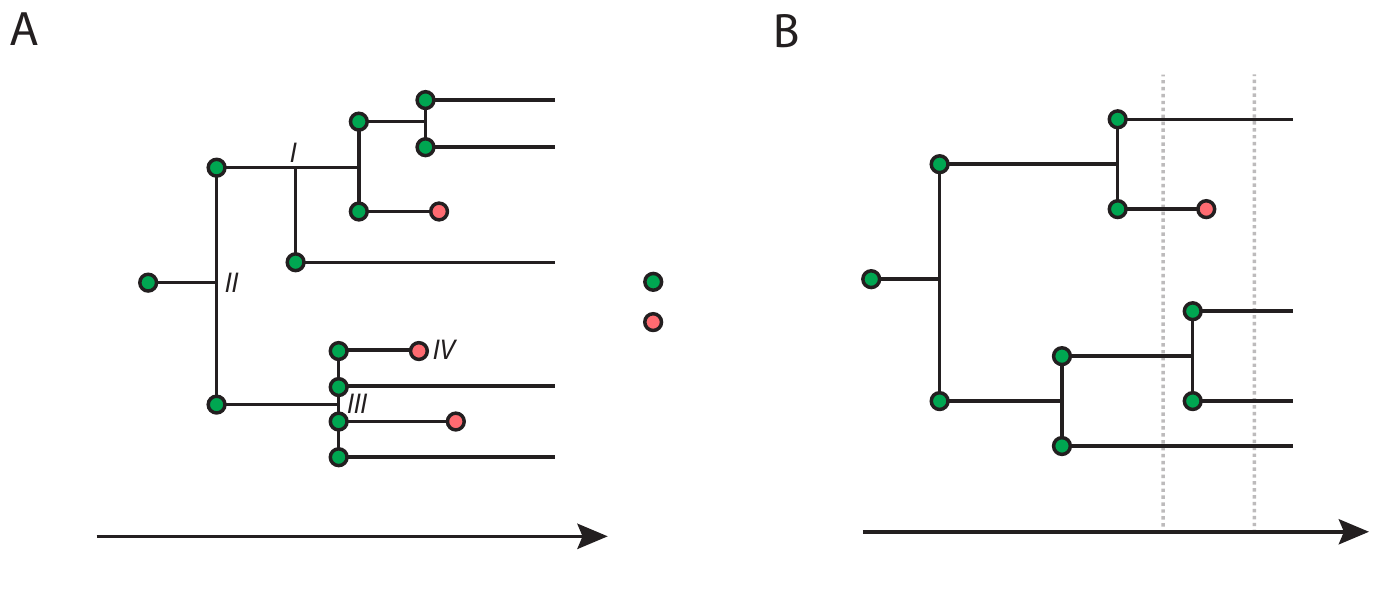}
\put(-20,0){Time}
\put(-37,5){$t_2$}
\put(-61,5){$t_1$}
\put(-226,0){Time}
\put(-190,79){Birth}
\put(-190,69){Death}
\caption{(A) A generalized branching process. At I we have a budding (or simple) mode of birth, where the parental individual continues to exist after birth. At II we have binary fission, where the parental individual is terminated at the moment two identical twin daughters are created. The number of progeny in fission can vary, such as at III where four daughters are born, or the death process at IV, which can be viewed as fission with no progeny. (B) A binary fission process typically seen in cell division. At time $t_1$ we have two sets of twins, each pair having identical ages. At time $t_2$ one has died and one has divided, leaving a single pair of twins, and two singletons with distinct ages.}
\label{F1_MicroMod}
\end{figure}

These can be modeled with branching processes, such as portrayed in Fig. \ref{F1_MicroMod}A, where we see a population growing through time, starting from a single individual. The differing nature of branching processes can be seen to significantly influence dynamic effects. At \emph{I}, for example, we have a \emph{simple} (or \emph{budding}) mode of \emph{birth}, such as that observed in yeast cell division \cite{Shcheprova2008, Lippuner2014}, where the parental cell survives the creation of an offspring cell. At \emph{II} we have \emph{binary fission}, more commonly associated with cell division, where the parental cell effectively terminates at the moment two identical twin daughter cells are created. Fig. \ref{F1_MicroMod}B contains a full example of this, which will later be examined in detail. At \emph{III} we see a more general fission process, with four daughter cells. Along with such birth processes, there are also \emph{death} processes, such as at \emph{IV}, which can also be viewed as fission with zero progeny.

Simpler models of branching processes generally assume dynamics only depend upon the current population size. However, the age of individuals is an important factor in population growth, with fecundity often a function of age. Cell division is most likely to occur after the cell has had sufficient time to mature and enter the mitosis phase, for example. In terms of Fig. \ref{F1_MicroMod}, incorporating age effects is equivalent to assuming the lengths of the branches affect dynamics. The incorporation of age into population dynamics was initially described by the McKendrick-von Foerster equation \cite{McKendrick1926, VonFoerster1959}, where a partial differential equation approach was used to describe the evolution of a continuous age structure. More recent developments utilize the machinery of semi-groups and have included other effects such as spatial dependence and cell division processes that depend upon the cells size. More details of these methods and their applications can be found in \cite{Webb2008}. Leslie matrices \cite{Leslie1945, Leslie1948} are a discretized version of the process modeled by the McKendrick-von Foerster equation. These linear models have certain algebraic advantages and have proved popular in ecological modeling \cite{Caswell2001}. However, these models ignore the stochastic fluctuations in population size. The Bellman-Harris process went some way in addressing this problem \cite{Bellman1948}, providing an integral equation for the sample size generating function. Although this has an implicit dependence on the underlying age dependence, solving this equation does not reveal the age-structure for the system. It is also designed for fission events (such as events \emph{II}, \emph{III} or \emph{IV} in Fig. \ref{F1_MicroMod}B) and does not cater for simple birth (e.g., event \emph{I}). More recent fully stochastic approaches have emerged that utilize Martingale techniques \cite{Jagers2000, Hamza2016, Hong2011, Hong2013} and adapt methods from gas kinetics \cite{ Greenman2016, Chou2016}.

Temporal fluctuations in population number have parallels with particle physics, where the number of subatomic particles can change due to interactions arising from fundamental forces. This parallel was first noted by Doi \cite{Doi1976,Doi1976b}, who adapted time-ordered perturbative expansion techniques to chemical reaction processes. An extensive review of this approach applied to a range of problems, including spatial effects, can be found in \cite{Mattis1998}. Peliti constructed a path integral formulation, developing lattice methods for birth death processes  \cite{Peliti1985}. Although these methods are generally bosonic, where lattice sites can have multiple occupation, they have also been extended for exclusive occupation (fermionic) processes \cite{Schulz2005}. A recent review of path integral techniques for master equations can be found in \cite{Weber2016}. Significant work in the development of renormalization techniques has also taken place \cite{Peliti1986, Lee1994, Tauber2005}. Recent pedagogic expositions can be found in \cite{Tauber2012, Tauber2014}. To date there has been no application of these methods to age dependent processes. In this work we address this, and show that Doi-Peliti methods offer an efficient approach to such problems.

In the next section we give an outline of age structured population dynamics. Section $3$ details how Doi-Peliti formalism can be adapted to age-structured problems. In particular, it is used to derive kinetic equations for correlation functions, and full age-structure-population-size probability densities. Section 4 derives path integral solutions to these equations, obtaining an exact perturbative expansion for correlation functions. This expansion can be directly related to the hereditary structure of correlated individuals and is our main result. Section $5$ generalizes these methods to binary fission models of cell population dynamics. Conclusions complete the study.

\section{Age-Structured Dynamics}

\subsection{Background}

The majority of methods analyzing the age structure of a population consider a distribution of the form $X(q,t)$, where $\int_\Omega X(q,t)\dd q$ represents the number of individuals from the population of interest with age in the set $\Omega$. The backbone of these modeling techniques is the McKendrick-von Foerster equation \cite{McKendrick1926, VonFoerster1959, Webb2008}

\begin{equation}
{\partial X(q,t)\over \partial t} + {\partial X(q,t)\over \partial q} = 
-\mu(q)X(q,t),
\label{MCKENDRICK0}
\end{equation}
where $\mu(q)$ represents the \emph{death rate} per individual of age $q$. In survival statistics this is also referred to as the \emph{hazard function}, where $\mu(q) \dd t$ represents the probability that an individual with current age $q$ will subsequently die within an infinitesimal amount of time $\dd t$. The death rate can also be written as $\mu(q) = \frac{m(q)}{1-M(q)}$, where $m(q)$ is the probability density function for the waiting time until death occurs, with cumulative density function $M(q)$. The \emph{survival} function can thus be written as $S(q)=1-M(q)=e^{-\int_0^q\mu(u)du}$, representing the probability that an individuals time of death is greater than $q$. More details of the relationships between these terms can be found in \cite{Collett2015}.

For the McKendrick-von Foerster equation there is also an associated boundary condition representing the arrival of newborn individuals of age $q=0$,

\begin{equation}
X(q=0,t) =  \int_{0}^{\infty}
\beta(u)X(u,t)\dd u,
\label{MCKENDRICKBC}
\end{equation}
where $\beta(q)$ represents the \emph{birth rate} per parental individual of age $q$. The birth rate $\beta(q)=\frac{b(q)}{1-B(q)}$ also has corresponding waiting time distributions $b(q)$ and $B(q)$. This formulation of the McKendrick-von Foerster equation is for the simple mode of birth, such as the case labelled \emph{I} in Fig. \ref{F1_MicroMod}. An initial condition $X(q,t=0) = g(q)$ describing the age distribution of founder individuals completely specifies the mathematical model.

However, from the McKendrick-von Foerster equation we find the total population $\int_0^\infty X(q,t)\dd q$ grows deterministically and the stochastic fluctuations in population size are not incorporated. The natural way to model such variation is with the standard forward continuous-time master equation \cite{VanKampen2011,Chou2014}:

\begin{equation}
{\partial \rho_{n}(t)\over \partial t} = 
-n\left[\beta_n(t)+\mu_n(t)\right]\rho_n(t)
+(n-1)\beta_{n-1}(t)\rho_{n-1}(t) 
+(n+1)\mu_{n+1}(t)\rho_{n+1}(t),
\label{MASTER0}
\end{equation}
where $\rho_n(t)$ is the probability of population size $n$ at time $t$, and $\beta_{n}(t)$ and $\mu_{n}(t)$ are the birth and death rates, per individual, respectively. Each of these rates can be population size and time dependent.

This master equation does not consider age effects, however, and methods that incorporate both stochastic fluctuations and age are required. The models to do this will differ depending on the microscopic behavior of the corresponding branching process. We next develop the theory for simple budding birth such as at event \emph{I} in Fig. \ref{F1_MicroMod}A. More complex modes of birth such as binary fission at event \emph{II} shall later be developed in Section \ref{FISSSEC}.    


\subsection{Simple Budding Birth and Death Processes}

We consider the following stochastic process describing a population of $n$ individuals at time $t$. Each individual of age $p$ is represented by $A(p)$. We have a death process $A(p) \rightarrow \phi$ occurring at rate $\mu_n(p)$ that can depend upon age $p$ and population size $n$. We also have a budding birth process $A(p) \rightarrow A(p) + A(0)$ giving rise to a new individual of age zero, along with the surviving parent, occurring at a rate $\beta_n(p)$.

To represent this process requires some terminology. The vector $\q_n=(q_1,q_2,\dots,q_n)$ denotes the ages of the population of size $n$. The term $\rho_n(\q_n;t)\dd \q_n$ represents the probability that the population is of size $n$ and, if the $n$ individuals are randomly labeled $1,2,\dots,n$, the $i^\textrm{th}$ individual has age in the interval $[q_i,q_i + \dd q_i]$, at time $t$. The probability density $\rho_n$ is thus a symmetric function of the age arguments $\q_n$. 
The marginal probability densities are defined as $\rho_n^{(m)}(\q_m;t)=\int \dd \q'_{n-m}\HS\rho_n(\q_m,\q'_{n-m};t)$.

We also introduce the term $f_n(\q_n;t)$ as the probability density for the case where the ages are ordered, $q_1 \le q_2 \le \dots \le q_n$. The age domain can be extended to $\mathbb{R}^n$ by defining $f_n(\q_n;t) = f_n(\pi(\q_n);t)$, where $\pi$ is the permutation that orders the components of the vector $\q_n$. Note that $f_n \equiv n!\rho_n$. With this new domain, $f_n$ is no longer a probability density, however, many expressions are simpler to express with $f_n$ and we shall interchange between the two functions. 

The final term of interest considered is the correlation functions $X_m(\q_m;t)$, where $X_m(\q_m;t)\dd \q_m = \sum_{n \ge m}(n)_m\rho_n^{(m)}(\q_m;t)$ represents the probability that we can find $m$ individuals in the population and label them $1,2,\dots,m$ such that the $i^\textrm{th}$ has age in $[q_i,q_i+dq_i]$. Here $(n)_m = n(n-1) \dots(n-(m-1))$ denotes the Pochhammer symbol.

\begin{figure}[t]
\hspace{10mm}\includegraphics[width = 12.9cm,center]{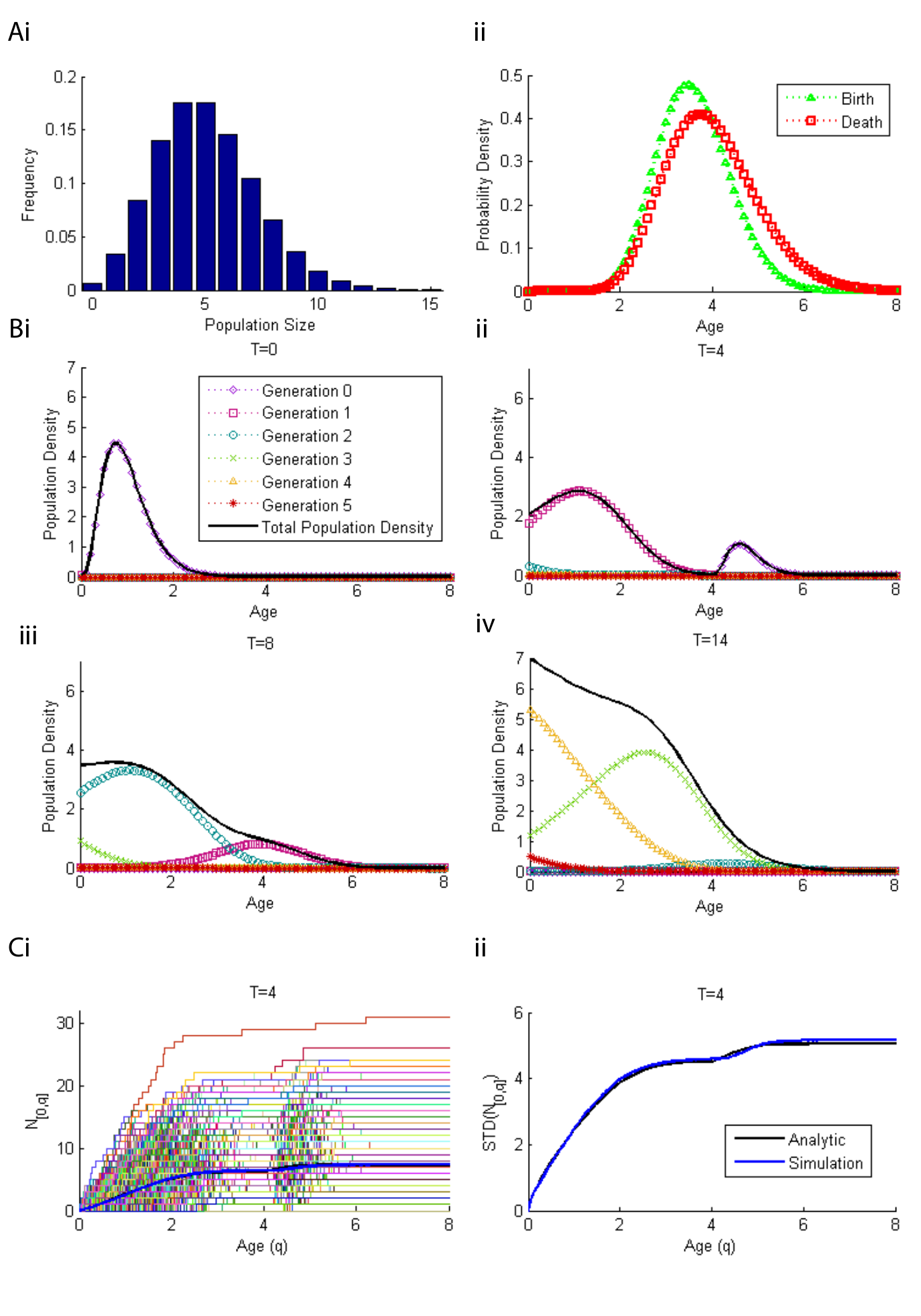}
\caption{An age structured budding birth death process. (Ai) Poisson distributed initial population size with rate parameter $5$. (Aii) Birth and death rates dependency upon age (see Section \ref{Numerics} for specifics). (Bi-iv) Population densities for a birth death process at times $T=0,4,8,14$. Distributions for individual generations are highlighted. (Ci) Plots of $N_{[0,q]}$ (the number of individuals younger than age $q$) at time $T=4$ are plotted for 1000 simulated birth death processes. Each color is one simulation. The two solid thick lines are the simulation mean (blue) and $\textrm{E}(N_{[0,q]})$ (black). (Cii) Standard deviation of simulations in Ci and $\textrm{STD}(N_{[0,q]})$ (see Eq. \ref{VarEq}).}
\label{QFT_Fig2}
\end{figure} 

More details about this representation can be found in \cite{Greenman2016, Chou2016}, where the probability density function $\rho_n(\q_n;t)$ and correlation function $X_n(\q_n;t)$ are examined in greater depth, which we summarize below. These results shall later be rederived using Doi-Peliti methods in Section \ref{DoiIntrSec}.

For the density $\rho_n(\q_n;t)$ the following hierarchy of equations was established:

\begin{eqnarray}
\displaystyle {\partial \rho_{n}(\q_{n};t)\over \partial t} + 
\sum_{j=1}^{n}{\partial \rho_{n}(\q_{n};t)\over \partial q_{j}} = 
-\rho_{n}(\q_{n};t) & \sum_{i=1}^{n} \gamma_{n}(q_{i}) \nonumber\\
 & + (n+1)\!\int_{0}^{\infty}\!\!\mu_{n+1}(y)
\rho_{n+1}(\q_{n},y;t)\dd y, 
\label{RHO0} 
\end{eqnarray}
along with boundary condition

\begin{equation}
n \rho_{n}(\q_{n-1},0;t) = \rho_{n-1}(\q_{n-1};t)\sum_{i=1}^{n-1}\beta_{n-1}(q_{i}).
\label{BCRHO}
\end{equation}
Note that the birth and death rates $\beta_n(q)$ and $\mu_n(q)$ can be functions of both population size $n$ and age $q$, and we have used the event rate $\gamma_n(q) = \beta_n(q) + \mu_n(q)$. This system was solved analytically in \cite{Greenman2016,Chou2016} for the cases of either pure birth ($\mu_n(q)=0$) or pure death ($\beta_n(q)=0$). Furthermore, provided the rates $\beta_n(q)=\beta_n$ and $\mu_n(q)=\mu_n$ are independent of age, the population size density $\rho_n^{(0)}(t)=\int \dd \q_n \rho_n(\q;t)$ was shown to reduce to master Eq. \ref{MASTER0}. 

For the $m^\textrm{th}$ order correlation function $X_m(\q_m;t)$, provided the birth and death rates rates $\beta_n(q)=\beta(q)$ and $\mu_n(q)=\mu(q)$ are independent of population size, the following equation was obtained:

\begin{equation}
\frac{\partial X_{m}}{\partial t} + \sum_{j=1}^{m}\frac{\partial X_{m}}{\partial q_{j}} + X_{m}\sum_{i=1}^{m} \mu_{n}(q_{i})  = 0,
\label{MCKGEN}
\end{equation}
with boundary condition

\begin{equation}
X_m(\q_{m-1},0;t) = X_{m-1}(\q_{m-1};t)\sum_{i=1}^{m-1}\beta(q_i)+\int_0^\infty X_m(\q_{m-1},y;t)\beta(y) \dd y.
\label{MCGENBC}
\end{equation}
For the mean-field case ($m=1$), this equation is precisely the McKendrick-von Foerster Eq. \ref{MCKENDRICK0}, and we find the correlation function satisfies a natural generalization.

Now, the hierarchy for the full probability densities given by Eq. \ref{RHO0} and \ref{BCRHO} express $\rho_n$ in terms of both lower and higher order terms $\rho_{n-1}$ and $\rho_{n+1}$. This mirrors the BBGKY hierarchies seen in gas kinetics \cite{McQuarrie2000, Zanette1990}, which are notoriously difficult to solve, and approximate closure schemes are often employed \cite{Raghib2011,Rogers2011}. However, we note that the hierarchy for the correlation functions given in Eq. \ref{MCKGEN} and \ref{MCGENBC} expresses $X_m$ just in terms of the lower order term $X_{m-1}$ suggesting that this hierarchy is easier to analyze (by back substitution for example, starting from the solution $X_1$ to the McKendrick-von Foerster equation). This analytic simplification is something that we will also see reflected in the path integral techniques applied in Section \ref{PIR}, where we will find $X_m$ easier to analyze than $\rho_n$.

Finally, we consider age-structured population-size variance. We first let $N_\Omega$ denote the random variable counting the number of individuals with age inside $\Omega$. We can then relate Eq. \ref{MCKGEN} to the variance $\textrm{Var}(N_\Omega)$ as follows.

\begin{eqnarray}
\textrm{Var}(N_\Omega) & = \textrm{Var}\left(\sum_{du \in \Omega}N_{du}\right)=\sum_{du,dv \in \Omega}\textrm{Cov}(N_{du},N_{dv})\nonumber\\
& =\sum_{du\in \Omega}\textrm{Var}(N_{du})+\sum_{du \ne dv \in \Omega}\textrm{Cov}(N_{du},N_{dv}),
\end{eqnarray}

Now, a sufficiently small interval $du$ will contain the age of at most one individual, and so $N_{du}$ is simply a Bernoulli variable indexing whether an individual exists in $du$. Thus $N_{du}^2=N_{du}$ and so $\textrm{E}(N_{du}^2)=\textrm{E}(N_{du})=X_1(u)du$. Also, $\textrm{E}(N_{du})^2$ is vanishingly small, and we find $\sum_{du\in \Omega}\textrm{Var}(N_{du})=\int_\Omega X_1(u) \dd u$. We similarly note that $\textrm{E}(N_{du}N_{dv})=X_2(u,v)\dd u\dd v$ is the probability of finding a pair of individuals with ages in intervals $du,dv$. Thus we finally obtain the following, correcting an error in \cite{Chou2016}:

\begin{equation}
\textrm{Var}(N_\Omega)  = \int_\Omega X_1(u) \dd u + \int_{\Omega^2} \left(X_2(u,v)-X_1(u)X_1(v)\right) \dd u \dd v.
\label{VarEq}
\end{equation}


\subsection{Numerics}
\label{Numerics}

We next exemplify the simple budding birth death process of Fig. \ref{QFT_Fig2} with a comparison of simulations to analytic expression for the mean-field and variance of the process. We suppose the death waiting time probability density function $m(q)=\Gamma(k)^{-1} \theta^{-k} q^{k-1}e^{-q/\theta}$ is gamma distributed with mean $4$ and variance $1$ ($k=16$, $\theta=\frac{1}{4}$). We suppose the birth rate $\beta(q)=c q^z\mu(q)$, where $c = 1.2$ and $z=0.2$. The resultant densities $b(q)$ and $m(q)$ are plotted in Fig. \ref{QFT_Fig2}Aii. Note that the birth rate is higher than the death rate and a mean increase in population is expected. We suppose the initial population size at time $t=0$ is Poisson distributed with mean $5$ (Fig. \ref{QFT_Fig2}Ai), where each founder individual has an age that is gamma distributed with mean $1$ and variance $\frac{1}{4}$ ($k=4$, $\theta=\frac{1}{4}$; Fig. \ref{QFT_Fig2}Bi). In Fig. \ref{QFT_Fig2}B we see the mean field solution $X_1(q;t)$ to the McKendrick-von Foerster equation. We have the age distribution of the generation zero founder individuals (\ref{QFT_Fig2}Bi). After a time $t=4$ (\ref{QFT_Fig2}Bii) this peak has significantly reduced due to death, and moved to the right, due to aging. The birth process has resulted in a new peak of individuals from the next generation. By time $t=14$ (\ref{QFT_Fig2}Biv)we see that the majority of individuals are from generations three and four, which significantly overlap in age. Note that the total population (area under black curve) has significantly increased from the initial  case $t=0$. We see from the simulations in \ref{QFT_Fig2}C that stochastic fluctuations lead to significant differences from the mean field, although the mean and variance of the simulations can be seen to match the solutions to Eq.s \ref{MCKENDRICK0} and \ref{VarEq} (\ref{QFT_Fig2}Cii).

This leaves open a two-fold problem. Firstly, how to solve Eq. \ref{MCKGEN} for the correlation functions $X_m(\q_m;t)$ (the boundary conditions in Eq. \ref{MCGENBC} complicate the derivation of an explicit solution; see \cite{Chou2016}). Secondly, how to derive the density $f_n(\q_n;t)$. To attack these problems we will use quantum field theoretic techniques, where we find that these two terms have very similar path integral formulations. In particular, we shall show that correlation functions restricted by generation numbers (such as the mean field densities in Fig. \ref{QFT_Fig2}B) and hereditary structure arise naturally in path integral perturbative expansions.


\section{Doi Peliti Second Quantization Methods}
\label{DoiIntrSec}

\subsection{Machinery}

Here, we introduce a Doi-Peliti operator formalism tailored to age-dependent birth-death processes. The use of field theoretic techniques was initiated by Doi \cite{Doi1976,Doi1976b}. This was based upon a \emph{state} vector $\ket{\q_n}\equiv\ket{q_1,q_2,\cdots,q_n}$ used to represent a set of $n$ objects, with corresponding properties $q_1,q_2,\cdots,q_n$, such as the coordinates of $n$ molecules for example. However, we interpret them as the ages of $n$ individuals. In this representation individuals are indistinguishable and the order of the components $q_i$ is immaterial. We use $\ket{\phi}$ to represent the `vacuum' state consisting of an empty population.

Next, we introduce annihilation and creation operators $\psi_q$ and $\psi_q^\dag$ which satisfy standard commutation relations:

\begin{equation}
[\psi_q,\psi_{p}^\dag] = \delta(q-p), \hspace{2mm}[\psi_q,\psi_{p}] =
[\psi_q^\dag,\psi_{p}^\dag]=0.
\label{CommRel}
\end{equation}
States can be constructed using creation operators: $\ket{\q_n}=
\psi_{q_1}^\dag\psi_{q_2}^\dag\cdots\psi_{q_n}^\dag\ket{\phi}$.  From
these states and commutation relations, we obtain the normalization

\begin{equation}
\braket{\q_m|\p_n}=\delta_{mn}\sum_{\pi \in S_m}\prod_{i=1}^n\delta(q_i-p_{\pi(i)}),
\label{NORMA}
\end{equation}
where $\braket{\phi|\phi}\equiv 1$ and $S_m$ is the symmetry group of permutations on $m$ symbols.  The annihilation operators are assumed to kill the vacuum state; $\psi_q\ket{\phi} = 0$.  Note that although we will use this formalism to model (positive) ages, we place no restrictions on the states $\ket{\q_n}$ which can contain negative entries. It is relatively straightforward to use Eq. \ref{NORMA} to verify the following resolution of the identity operator:

\begin{equation}
I = \sum_{m=0}^\infty\int \frac{\dd q_m}{m!}\ket{\q_m}\bra{\q_m}.
\label{RESID}
\end{equation}

Next, probabilty is introduced into the system by defining the following \emph{superposition} of states:

\begin{equation}
\ket{f(t)}=\sum_{n=0}^\infty \int_{\mathbb{R}^n}\frac{\dd \q_n}{n!}f_n(\q_n;t)\ket{\q_n}.
\label{SUPPOS}
\end{equation}
This is the representation used in Doi \cite{Doi1976,Doi1976b}, where the distribution $f_n(\q_n;t) \equiv n!\rho_n(\q_n;t)$ is equivalent to the one mentioned in the previous section. The utility of this representation is manifest when the evolution of the state can be described linearly as,

\begin{equation}
\frac{\partial}{\partial t}\ket{f(t)}=\zeta\ket{f(t)},
\label{TimeEvolEq}
\end{equation}
with a suitable operator $\zeta$. For our application  $\zeta= \zeta_0+\zeta_b+\zeta_d$ can be decomposed into three parts. The term $\zeta_0$ describes the increase of all age variables in time, the term $\zeta_b$ represents the increase in population size due to birth and $\zeta_d$ represents the decrease in population size due to death. Then we have, following Doi methodology \cite{Doi1976,Doi1976b}, the following expressions:

\begin{equation}
\zeta_0 = \int \dd q \hspace{1mm} \psi_q^\dag \frac{\partial}{\partial q}\psi_q,\hspace{4mm}
\zeta_b = \int \dd q \hspace{1mm} \beta(q)(\psi_q^\dag\psi_q-\psi_q^\dag\psi_0^\dag\psi_q),\hspace{4mm}
\zeta_d = \int \dd q \hspace{1mm} \mu(q)(\psi_q^\dag\psi_q-\psi_q).
\label{MICRODEF}
\end{equation}
Thus for example, the second term in $\zeta_b$ contains creation operator $\psi^\dag_0$, representing the birth of a new individual of age zero, and the annihilation and creation operators $\psi_q$ and $\psi_q^\dag$ are a bookkeeping measure that preserves the parental individual of age $q$. These states and operators are given in the \emph{Schr{\"o}dinger} representation where the operators are constant and the states vary in time, with Eq. \ref{TimeEvolEq} having a formal solution of the form

\begin{equation}
\ket{f(t)}=e^{-\zeta t}\ket{f(0)}.
\end{equation}

We complete this section by introducing functional \emph{coherent} states. If we take any complex function $u(q)$ of the real value $q$, with complex conjugate $u^*(q)$, we can construct the the following state superposition:

\begin{equation}
\ket{u} = e^{\int dq \hspace{1mm} u(q) \psi_q^\dag}\ket{\phi}
=\sum_{n=0}^\infty\int \frac{\dd \q_n}{n!} 
u(q_1)\cdots u(q_n) \psi_{q_1}^\dag\cdots
\psi_{q_n}^\dag\ket{\phi}.
\label{COHE}
\end{equation}
These coherent states satisfy the following eigenstate property:

\begin{eqnarray}
\psi_p\ket{u} & = \sum_{n=0}^\infty \frac{1}{n!}\int \dd \q_n \hspace{1mm} u(q_1)\cdots u(q_n)\psi_p\psi_{q_1}^\dag\cdots\psi_{q_n}^\dag\ket{\phi}\nonumber\\
& =  \sum_{n=0}^\infty \frac{1}{n!}\int \dd \q_n \hspace{1mm} u(q_1)\cdots u(q_n)\sum_{i=1}^n\delta(p-q_i)\prod_{j \ne i}\psi_{q_j}^\dag\ket{\phi} =  u(p)\ket{u}.
\label{EIG}
\end{eqnarray}

Coherent states also satisfy the following normalization property, 

\begin{equation}
\braket{u|u'} = \bra{\phi}e^{\int dq \HS u^*\psi_q}e^{\int dq\HS u'\psi_q^\dag}\ket{\phi} = \bra{\phi}e^{\int dq \HS u^*u'} 
e^{\int dq\HS u'\psi_q^\dag}e^{\int dq \HS u^*\psi_q}\ket{\phi} = e^{\int dq\HS u^*u'},
\label{CohNorm}
\end{equation}
where we have used the Baker-Campbell-Hausdorff theorem to commute operators \cite{Maggiore2004, Peskin1995}.

The functional coherent states $\ket{u(q)}$ generalize the constant coherent states $\ket{z}$ used by Doi \cite{Doi1976, Doi1976b}, which can be recovered by setting the function $u(q)\equiv z$ to be constant. Doi noticed the form $\ket{z}$ allows many summary statistics of interest to be simply expressed. In particular,  the correlation density $X_m(\q_m;t)$ is given by the following expectation of the number operator $\psi_{q_1}^\dag\cdots \psi_{q_n}^\dag\psi_{q_1} \cdots\psi_{q_n}$:

\begin{equation}
X_m(\q_m;t) = \braket{1|\psi_{q_1}^\dag\cdots\psi_{q_m}^\dag\psi_{q_1}\cdots\psi_{q_m}|f(t)}=\braket{1|\psi_{q_1}\cdots\psi_{q_m}e^{-\zeta t}|f(0)}.
\label{DENSDEF}
\end{equation}
Note that in this expression $\bra{1}$ represents a coherent state with function $u(q)\equiv1$ (rather than a state representing a single individual of age $1$). We have also used the fact that $\bra{1}$ is a left eigenstate of $\psi_{q_i}^\dag$ with eigenvalue $1$, as seen in Eq. \ref{EIG}.

Peliti \cite{Peliti1985} adapted these methods for birth death processes, developing path integral techniques rather than the time-ordered perturbative methods of Doi \cite{Doi1976, Doi1976b}. In particular, a state vector $\ket{n}$ was used to represent a population of $n$ individuals, and a functional mapping $\ket{n} \equiv z^n$ utilized to construct a path integral formulation for properties of interest, with lattice methods used to incorporate spatial effects. The more general coherent states we have introduced will allow us to directly construct a path integral formulation for our problem without recourse to lattice discretization methods.


\subsection{Kinetic Equations}
\label{KINSEC}

We now use the formalism outlined above to derive equations for the distribution $f_n(\q_n;t)$. Specifically, we utilize the field theoretic framework to derive the hierarchy given in Eq.s \ref{RHO0} and \ref{BCRHO}. If we condition upon an initial distribution $f_n(\q_n;0)$, then from Eq.s \ref{NORMA} and \ref{SUPPOS} we find that $f_n(\q_n;t)$ is the projection of the state $\ket{f(t)}$ onto the fundamental state $\ket{\q_n}$:

\begin{equation}
f_n(\q_n;t) = \braket{\q_n|f(t)} = \braket{\q_n|e^{-\zeta t}|f(0)}.
\label{FDEF}
\end{equation}
Since both $\bra{\q_n}$ and $\ket{f(0)}$ are constant in time, we can differentiate Eq. \ref{FDEF} with respect to time to find

\begin{equation}
\frac{\partial f_n(\q_n;t)}{\partial t} = -\braket{\q_n|\zeta|f(t)} =
-\braket{\q_n|\zeta_0|f(t)}-\braket{\q_n|\zeta_b|f(t)}
-\braket{\q_n|\zeta_d|f(t)}.
\label{MATEQ}
\end{equation}

Next we use the commutation relations to calculate the left action of the operators $\zeta_0, \zeta_b$ and $\zeta_d$ upon $\bra{\q_n}$. The first term gives

\begin{eqnarray}
\braket{\q_n|\zeta_0|f(t)} & = \int \dd p \braket{\phi|\psi_{q_1}\psi_{q_2}\cdots\psi_{q_n}\psi_p^\dag\frac{\partial}{\partial p}\psi_p|f(t)}
= \int \dd p \braket{\phi|\sum_{i=1}^n\delta(p-q_i)\prod_{j \ne i}\psi_{q_j}\frac{\partial}{\partial p}\psi_p|f(t)} 
 \nonumber\\
& = \sum_{i=1}^n\frac{\partial}{\partial q_i}\braket{\phi|\prod_{j \ne i}\psi_{q_j}\psi_{q_i}|f(t)}
= \sum_{i=1}^n\frac{\partial}{\partial q_i}\braket{\q_n|f(t)}
= \sum_{i=1}^n\frac{\partial}{\partial q_i}f_n(\q_n;t).
\end{eqnarray}
For the birth term, a similar derivation yields

\begin{eqnarray}
\braket{\q_n|\zeta_b|f(t)} & = \int \dd p \hspace{1mm} \beta(p)\braket{\phi|\psi_{q_1}\psi_{q_2}\cdots\psi_{q_n}(\psi_p^\dag\psi_p-\psi_p^\dag\psi_0^\dag\psi_p)|f(t)} 
\nonumber\\
& = \sum_{i=1}^n\beta(q_i)\left[f_n(\q_n;t)-\sum_{j \ne i}\delta(q_j)f_{n-1}(\q_n^{(-j)};t)\right].
\label{BIRTHDELTA}
\end{eqnarray}
where $\q_n^{(-j)}=(q_1,\cdots,q_{j-1},q_{j+1},\cdots,q_n)$ represents the age-chart with all $n$ ages except the $j^{th}$ one, which is omitted. Finally, the death term yields

\begin{equation}
\braket{\q_n|\zeta_d|f(t)} = \sum_{i=1}^n\mu(q_i)f_n(\q_n;t)-\int \dd p \hspace{1mm} \mu(p)f_{n+1}(\q_n,p;t).
\end{equation}

Upon combining these results for any strictly positive age chart $\q_n$ we lose the delta functions in Eq. \ref{BIRTHDELTA} and (using $\rho_n \equiv f_n/n!$) rediscover Eq. \ref{RHO0}. Furthermore, integrating Eq. \ref{MATEQ} with respect to the variable $q_j$ over a small interval containing the boundary $q_j=0$ captures a delta function from Eq. \ref{BIRTHDELTA} and recovers the boundary condition in Eq. \ref{BCRHO}.

The field theoretic methods thus provide an efficient means of deriving kinetic equations for the population distribution function $f_n(\q_n;t)$, and thus, a complete stochastic description of the population size and age structure for the entire population.


\subsection{Age-Structured Moments}

We now derive kinetic equations for the age structured correlation densities $X_m(\q_m;t)$  of the age-structure. Consider first the case $X_1(q;t) \equiv X(q;t)$. Then differentiating Eq. \ref{DENSDEF} with respect to time, with $m=1$, we find

\begin{equation}
\frac{\partial X}{\partial t} = -\braket{1|\psi_q\zeta|f(t)}=-\braket{1|\psi_q(\zeta_0+\zeta_b+\zeta_d)|f(t)}.
\end{equation}
This yields three terms on the right-hand side that can be written in the forms

\begin{eqnarray}
\fl
\braket{1|\psi_q\zeta_0|f(t)} & = \braket{1|\psi_q\int \dd p \psi_p^\dag\frac{\partial}{\partial p}\psi_p|f(t)}
 =  \int \dd p\HS \braket{1|\delta(q-p)\frac{\partial}{\partial p}\psi_p|f(t)}+\int \dd p\HS \braket{1|\psi_p^\dag\psi_q\frac{\partial}{\partial p}\psi_p|f(t)}
\nonumber\\
& =  \frac{\partial}{\partial q}\braket{1|\psi_q|f(t)}+\int \dd p\HS \frac{\partial}{\partial p}\braket{1|\psi_q\psi_p|f(t)} =  \frac{\partial X(q)}{\partial q} + \int \dd p\HS\frac{\partial}{\partial p}X_2(q,p), 
\label{TimeTerm}
\end{eqnarray}
\begin{eqnarray}
\fl
\braket{1|\psi_q\zeta_b|f(t)}& =  \braket{1|\psi_q\int \dd p\HS \beta(p)\left[\psi_p^\dag\psi_p-\psi_p^\dag\psi_0^\dag\psi_p\right]|f(t)}
\nonumber = \beta(q) X(q) + \int \dd p\HS\beta(p)\braket{1|\psi_q\psi_p|f(t)}
\nonumber\\
& \hspace{5mm}-\beta(q) X(q)-\delta(q)\int \dd p \HS\beta(p)\braket{1|\psi_p|f(t)}-\int \dd p\HS\beta(p)\braket{1|\psi_q\psi_p|f(t)}
\nonumber\\
& =  -\delta(q)\int \dd p\HS\beta(p)X(p),
\end{eqnarray}
and
\begin{eqnarray}
\fl
\braket{1|\psi_q\zeta_d|f(t)} & = \braket{1|\psi_q\int \dd p\HS \mu(p)\left[\psi_p^\dag\psi_p-\psi_p\right]|f(t)}
\nonumber\\
& = \mu(q) X(q) + \int \dd p\HS\mu(p)\braket{1|\psi_q\psi_p|f(t)}-\int \dd p\HS\mu(p)\braket{1|\psi_q\psi_p|f(t)} = \mu(q)X(q).
\end{eqnarray}
Then assuming $X_2(q,p)$ vanishes for the extreme values of $p$, the last term in Eq. \ref{TimeTerm} vanishes, yielding the PDE

\begin{equation}
\frac{\partial X}{\partial t} + \frac{\partial X}{\partial q} + \mu(q)X=\delta(q)\int \dd p\HS\beta(p)X(p).
\label{McKendrickInOne}
\end{equation}
For any $q>0$ we lose the delta function and recover the McKendrick-von Foerster Eq. \ref{MCKENDRICK0}, as expected. If we integrate across a vanishing small interval containing $q=0$, we also recover the McKendrick-von Foerster boundary condition of Eq. \ref{MCKENDRICKBC}.

Higher order correlations $X_m(\q_m;t)$ obey the following equation, which can be derived in much the same way as Eq. \ref{McKendrickInOne}; by differentiating Eq. \ref{DENSDEF} with respect to time and calculating the resulting matrix elements, the details of which are left to the reader:

\begin{equation}
\fl
\frac{\partial X_m}{\partial t} + \sum_{i=1}^m\frac{\partial X_m}{\partial q_i} + X_m\sum_{i=1}^m\mu(q_i)=\sum_{i=1}^n\delta(q_i)\left[\int \dd p\HS\beta(p)X_m(\q_m^{(-i)},p;t)+X_{m-1}(\q_m^{(-i)};t)\sum_{j \ne i}\beta(q_j)\right].
\label{McKGENN}
\end{equation}
This equation is equivalent to Eq. \ref{MCKGEN} and its associated boundary condition, and we find field theoretic methods provide a natural way to describe the correlation functions for age-structured populations.


\section{Path Integral Representation}
\label{PIR}

The path integral formulation of quantum mechanics works well in part because the fundamental position and momentum states are eigenstates of terms in Hamiltonians corresponding to many fields of interest. These fundamental states are then use to construct resolutions of the identity, which are applied between time slices across the time period of interest, resulting in path integrals \cite{Sredniki2007, Feynman1965}. This technique will not work for the systems we consider. Specifically, we have fundamental states $\ket{\q_n}$ that represent the age-charts of populations of size $n$. However, the `Hamiltonians' $\zeta$ we consider are functions of creation and annihilation operators, and the states $\ket{\q_n}$ are not eigenstates for these operators. Creation operators increase the minimum occupation number for any state superposition indicating that an eigenstate will not exist. However, eigenstates exist for annihilation operators; the coherent states we have seen in Eq. \ref{EIG}. To construct a path integral, we thus need a resolution of the identity in terms of coherent states. There are two possible approaches. One generalizes that of Peliti \cite{Peliti1985} and is the approach we take, as detailed in Appendix A. The other approach adapts techniques more commonly found in quantum field theoretic applications, as detailed in Appendix B, along with an explanation why two path integral formulations are possible.


\subsection{Construction}

We have, then, the following path integral resolution of the identity,

\begin{equation}
I = \int\mathcal{D}u\mathcal{D}v\HS e^{-i\int
dq\hspace{0.5mm} u(q)v(q)}\ket{iv}\bra{u}, 
\label{ResId}
\end{equation}
where the functional integrals over $u$ and $v$ are over real functions such that

\begin{equation}
\int\mathcal{D}u\mathcal{D}v\equiv \lim_{\scriptsize{\begin{array}{c}\epsilon\rightarrow 0\\Q\rightarrow \infty\end{array}}}
\prod_{\scriptsize{\begin{array}{c}q=-Q\\ \Delta q =\epsilon\end{array}}}^Q\int \frac{\epsilon}{2\pi}\dd[u(q)]\dd[v(q)].
\end{equation}

We next construct a path integral representation of the amplitude between two coherent states, using the resolution of the identity from Eq. \ref{ResId} at the time slices. To do this we first obtain matrix elements for the evolution operators $\zeta_0$, $\zeta_b$ and $\zeta_d$ between coherent states $\bra{u}$ and $\ket{iv}$. We find, using the eigenvalue properties of Eq. \ref{EIG}, that

\begin{eqnarray}
\braket{u|\zeta_0|iv} & = \braket{u|iv}i\int \dd q \HS u\frac{\partial v}{\partial q}\label{OperEx1},\nonumber\\
\braket{u|\zeta_b|iv} & = \braket{u|iv}i(1-u(0))\int \dd q\HS \beta uv\label{OperEx2},\nonumber\\
\braket{u|\zeta_d|iv} & = \braket{u|iv}i\int \dd q \HS \mu v(u-1) \label{OperEx3}.
\end{eqnarray}
For small time interval $\epsilon$, and using the normalization property of Eq. \ref{CohNorm}, we thus find

\begin{equation}
\braket{u|e^{-\zeta \epsilon}|iv} = \exp\left\{-\epsilon i \int\dd q \left[ -\frac{uv}{\epsilon} + u\frac{\partial v}{\partial q}+ (1-u(0))\beta uv +\mu v(u-1)\right]\right\}.
\end{equation}
Taking a product of such time slices over a time interval $[0,T]$, we obtain the following path integral formulation, where we have used integration by parts on the time differential term:

\begin{eqnarray}
\fl
\braket{u_T|e^{-\zeta T}|iv_0} & = \int \mathcal{D}u\mathcal{D}v
\exp\left\{-i\int \dd q \dd t\left[u\left(\frac{\partial v}{\partial q}+\frac{\partial v}{\partial t}\right)+\mu v(u-1)
+\beta uv\HS (1-u^0)\right] +i\int \dd q \hspace{1mm}u_T v_T\right\}.
\label{MatPI}
\end{eqnarray}
Here $u(q,t)$ and $v(q,t)$ are now real functions of age and time, and we introduce shorthand notation $u^0=u(0,t)$, $u_0=u(q,0)$, $u_T=u(q,T)$, $v_0=v(q,0)$ and $v_T=v(q,T)$. 

Now the two features of interest we investigate with this framework are the correlation density function $X_m(\q_m;T) = \braket{1|\psi_{q_1}\dots\psi_{q_m}e^{-\zeta T}|f(0)}$ and the density function $f_m(\q_m;t) = \braket{\q_m|e^{-\zeta T}|f(0)}=\braket{\phi|\psi_{q_1}\dots\psi_{q_m}e^{-\zeta T}|f(0)}$. In both cases we need to specify the initial population distribution state $\ket{f(0)}$. The formalism is most compact when this is a coherent state. Specifically, we assume that the initial population size is Poisson distributed with parameter $\alpha$, such that each individual has an initial age distribution $\omega(q)$. We then find that $\ket{f(0)}=e^{-\alpha}\ket{\alpha\omega}$.

Now using Eqs \ref{COHE} and \ref{CohNorm}, we obtain the following identities

\begin{equation}
\braket{1|iv_T} = e^{i\int dq\HS v_T},\hspace{2mm} \hspace{2mm}
\braket{u_0|f(0)} = e^{\alpha\int dq \HS \omega (u_0-1)} \hspace{2mm}
\textrm{and} \hspace{2mm}
\braket{\phi|iv_T}=1.
\label{BOUNDINFO}
\end{equation}
Then using resolutions of the identity at time points $0$ and $T$, placed at the right and left sides of operator $e^{-\zeta T}$ in Eq. \ref{MatPI}, respectively, we obtain the following expressions, where we have utilized the boundary information in Eq. \ref{BOUNDINFO}, the eigenstate property $\psi_{q_1}\dots \psi_{q_m}\ket{iv_T} =\prod_{j=1}^m\left[iv_T(q_j)\right]\ket{iv_T}$, and finally the field shift $u \rightarrow u+1$ to transform the integrals, giving:

\begin{eqnarray}
f_m(\q_m;T) =\int \mathcal{D}u\mathcal{D}v\prod_{j=1}^m
\left[iv_T(q_j)\right]& \exp\left\{-i\int \dd q \dd t \left[u\left(\frac{\partial v}{\partial q}+\frac{\partial v}{\partial t}+\mu v\right)
-\beta (u+1) u^0v \right]\right\}\cdot\nonumber\\
& \hspace{-15mm}\exp\left\{ - i\int \dd q \HS u_0 v_0 + \alpha\int \dd q \HS \omega u_0- i\int \dd q \HS v_T\right\},
\label{DENSPI}
\end{eqnarray}
and
\begin{eqnarray}
X_m(\q_m;T) =\int \mathcal{D}u\mathcal{D}v\prod_{j=1}^m
\left[iv_T(q_j)\right]& \exp\left\{-i\int \dd q \dd t \left[u\left(\frac{\partial v}{\partial q}+\frac{\partial v}{\partial t}+\mu v\right)
-\beta (u+1) u^0v \right]\right\}\cdot\nonumber\\
& \hspace{-15mm}\exp\left\{ - i\int \dd q \HS u_0 v_0 + \alpha\int \dd q \HS \omega u_0\right\}.
\label{MOMPI}
\end{eqnarray}

Although the two expressions are almost identical and both $f_m(\q_m; t)$ and $X_m(\q_m; t)$ can be examined by perturbative methods, it is only the latter expansion that readily combines into a simpler form.


\subsection{Perturbative Expansion}

The path integral in Eq. \ref{MOMPI} contains two interaction terms, the quadratic death term $\mu uv$ and the cubic birth term $\beta v(u+1) u(0,t)$. We calculate this by expanding the birth term and the initial term $\alpha \int \dd q \HS \omega u_0$. The death term is quadratic and so can be dealt with directly. We are then in a position to use standard methods, and require the following propagator $G(p,\tau;p',\tau')$, which we derive using generating functional techniques (see \cite{Mandl2010} for more discussion of such methods):

\begin{eqnarray}
\hspace{-10mm} G(p,\tau;p',\tau') & = \int \mathcal{D}u\mathcal{D}v \HS u(p,\tau)iv(p',\tau') \exp\left\{-i\int \dd q \dd t \left[u\left(\frac{\partial v}{\partial q}+\frac{\partial v}{\partial t}+\mu v \right) \right] - i\int \dd q \HS u_0 v_0\right\}\nonumber\\
& = \left.\frac{i\delta^2Z(J,K)}{\delta J(p;\tau)\delta K(p';\tau')} \right|_{J\equiv K\equiv 0},
\label{FDS}
\end{eqnarray}
where we introduce indeterminate functions $J=J(q,t)$ and $K=K(q,t)$ along with generating functional
\begin{eqnarray}
Z(J,K) = &\int \mathcal{D}u\mathcal{D}v \exp\left\{-i\int \dd q \HS\dd t \left[u\left(\frac{\partial v}{\partial q}+\frac{\partial v}{\partial t}+\mu v+J \right) \right]\right\}\nonumber\cdot\\
& \hspace{40mm}\exp\left\{- i\int \dd q \HS u_0 v_0 + i \int \dd q \HS\dd t \HS vK\right\}.
\end{eqnarray}

Now the functional integral over variable $u$ forces the constraint $\frac{\partial v}{\partial q}+\frac{\partial v}{\partial t}+\mu v + J =0$ and functionally integrating over $u_0$ forces the constraint $v_0=0$, which we solve to give,

\begin{equation}
v(q,t) = -\int_0^{t}J(q-t+s,s)e^{-\int_s^t\mu(q-t+x)dx}\dd s,
\label{UDEF}
\end{equation}
so the generating functional is given by
\begin{equation}
Z(J,K) = \exp\left\{ i\int \dd q \HS\dd t \HS v(q,t)K(q,t) \right\}.
\end{equation}

Performing the functional differentiations of Eq. \ref{FDS} then results in propagator

\begin{equation}
G(p,\tau;p',\tau') = \delta((p'-\tau')-(p-\tau))S(p,p')\theta(\tau'-\tau).
\end{equation}
The delta function in this expression ensures an individual of age $p$ at time $\tau$ has age $p'$ at time $\tau'$, the survival function $S(p,p')=\exp\left\{-\int_p^{p'}\mu(x)\dd x\right\}$ ensures the individual does not die between the two times, and the theta function ensures end point $\tau'$ is later then the initial time $\tau$ of the interval being propagated across.  

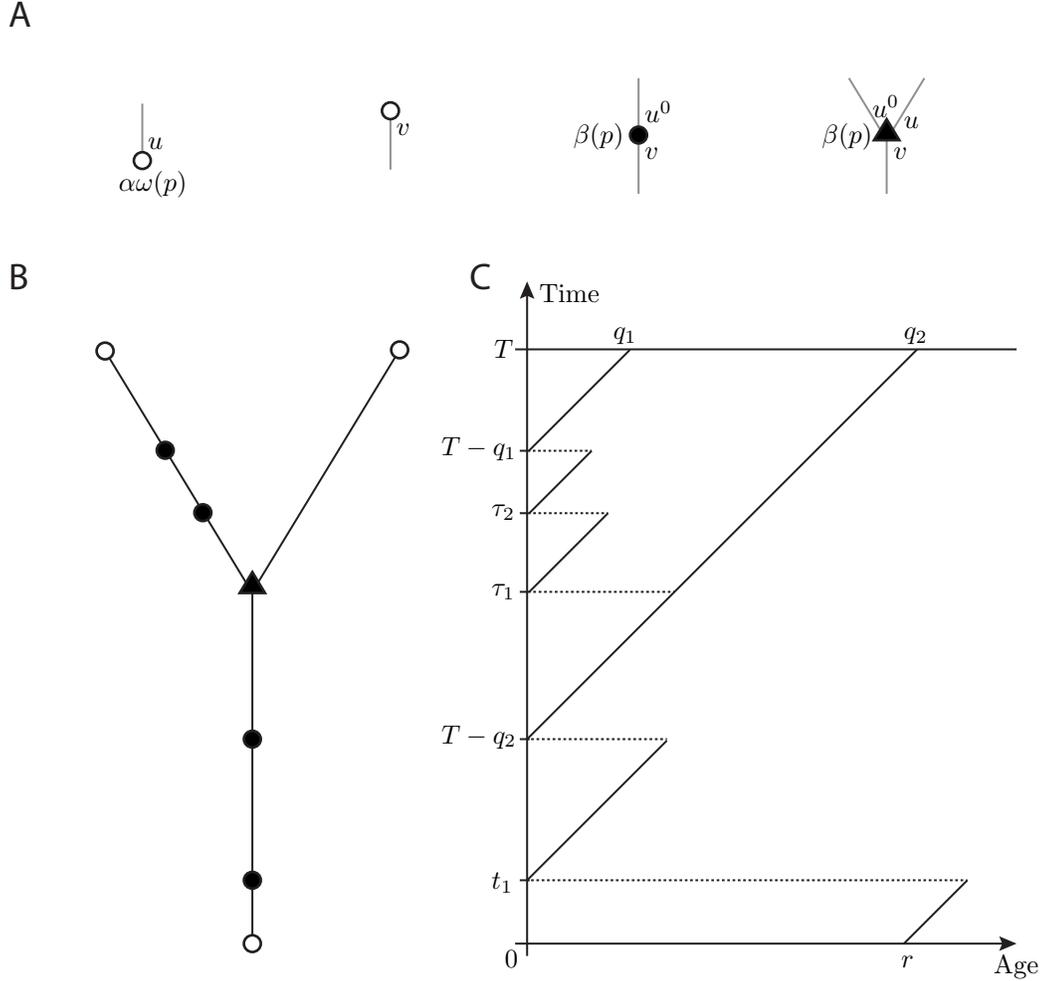
\begin{figure}[t]
\begin{center}
\includegraphics[height = 13cm]{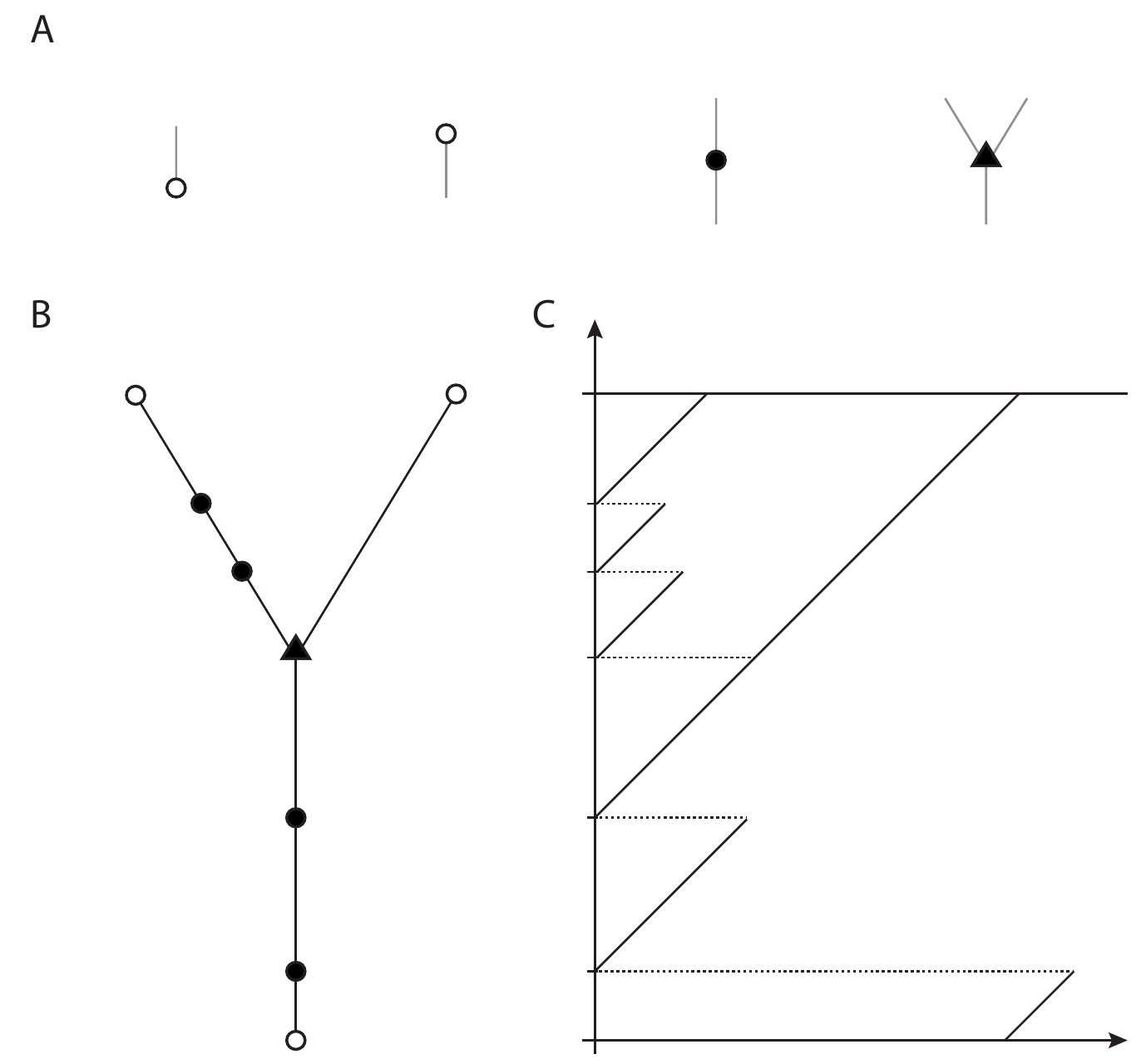}
\put(-15,-3){Age}
\put(-187,251){Time}
\put(-159,237){$q_1$}
\put(-49,237){$q_2$}
\put(-204,230){$T$}
\put(-224,193){$T-q_1$}
\put(-205,170){$\tau_2$}
\put(-205,140){$\tau_1$}
\put(-224,84){$T-q_2$}
\put(-205,29){$t_1$}
\put(-50,-1){$r$}
\put(-200,-1){$0$}
\put(-147,305){$v$}
\put(-147,319){$u^0$}
\put(-174,311){$\beta(p)$}
\put(-53,305){$v$}
\put(-49,317){$u$}
\put(-61,321){$u^0$}
\put(-80,311){$\beta(p)$}
\put(-241,314){$v$}
\put(-335,309){$u$}
\put(-346,293){$\alpha\omega(p)$}
\caption{Feynman diagram and associated Age-Time diagram. (A) Classes of node in Feynman diagram; one initiating, one terminating, and two internal nodes. (B) Feynman diagram consisting of one initiating node, two terminating nodes, four degree two nodes and one degree three node. (C) The associated Age-Time diagram. Individuals of age $q_1$ are generation five individuals, descending from generation two individuals of age $q_2$.}
\label{Hered_Exam}
\end{center}
\end{figure}

Finally we observe that the birth term $\beta (u+1)u^0v$ can be decomposed into two possible terms. Firstly, we can have $\beta uu^0v$, where the terminal variable $v$ at the end of one propagator gives rise to variables $u$ (representing parental individuals) and $u^0$ (representing new born individuals) at the beginning of two propagators. Alternatively, we can have term $\beta u^0v$, meaning we only get a single new propagator with initiating variable $u^0$. This means that there are two possible types of internal vertex in the corresponding Feynman diagram, of degrees three and two, respectively, as indicated in Fig. \ref{Hered_Exam}A.

In summary, then, to any Feynman diagram associated with the expansion of $X_m(\q_m;t)$, we associate the following. We have a certain number of initiating nodes, each associated with variable $\alpha\omega$. We have internal nodes from the two types mentioned above, each associated with a birth rate function $\beta(p)$, where $p$ is the age of the parent giving birth. We have $m$ terminating nodes. Propagators bridging the nodes then complete the picture. More explicit details of the terms and resulting integral can be found in Appendix E. 

These integrals can be simplified with the aid of Laplace transforms, as we now demonstrate with an example.Consider the pair correlation example in Fig. \ref{Hered_Exam}. Here we have a Feynman diagram (\ref{Hered_Exam}B) containing five birth terms, occurring at times $t_1$, $T-q_2$, $\tau_1$, $\tau_2$ and $T-q_1$, and one initiating event at age $r$. We also have a corresponding Age-Time picture (\ref{Hered_Exam}C). Although the Feynman diagram is a connected graph, the Age-Time representation is not; any propagated edge associated with a variable $u^0$ is representing a newborn individual, and the difference in age between parent and offspring results in a corresponding discontinuity. However, the advantage of this representation is that we can now see the hereditary structure between correlated individuals; the pair correlation in Fig. \ref{Hered_Exam}C is between ages of individuals in generations five and two, such that the generation five individuals are descendants of generation two individuals. This leaves us with an integral of the following form, where we define $Y(q)=\beta(q)S(q)$, with $S(q)$ shorthand for $S(0,q)$:

\begin{eqnarray}
X^{(2,0,3)}_2(q_1,q_2;T) = & \alpha S(q_1)S(q_2)\int\displaylimits_0^\infty \hspace{-1mm}\dd r\HS\frac{\omega(r)}{S(r)}
\int\displaylimits_0^{T-q_2} \hspace{-2mm}\dd t_1\HS Y(r+t_1)Y(T-q_2-t_1)\cdot\nonumber\\
& \int\displaylimits_{T-q_2<\tau_1<\tau_2<T-q_1}\hspace{-10mm} \dd \tau_1 \dd \tau_2 \HS \beta(\tau_1-[T-q_1])Y(\tau_2-\tau_1)Y([T-q_1]-\tau_2)
\end{eqnarray}

Now both the $t_1$ and $\tau_1,\tau_2$ integrals are convolutions, so can be calculated with the aid of Laplace transformations ($\mathcal{L}$) and their inverses ($\mathcal{L}_t^{-1}$) to give the following:

\begin{equation}
X^{(2,0,3)}_2(q_1,q_2) = \alpha S(q_1)S(q_2)\mathcal{L}_{q_2-q_1}^{-1}(\mathcal{L}(\beta)\mathcal{L}(Y)^2)
\int\displaylimits_0^\infty \hspace{-1mm}\dd r\HS\frac{\omega(r)}{S(r)}
\mathcal{L}_{T-q_2}^{-1} (\mathcal{L}(Y(t+r))\mathcal{L}(Y))
\label{HEREX}
\end{equation}

Thus we have a pair correlation function corresponding to a specific hereditary structure (the superscript of $X^{(2,0,3)}_2(q_1,q_2;T)$ labels the hereditary structure, which is explained in Appendix C). If the larger generation number is modified, say from five to seven, this is equivalent to adding another two birth events between times $T-q_2$ and $T-q_1$ in Fig. \ref{Hered_Exam}C. This will induce the change of $\mathcal{L}(Y)^2$ to $\mathcal{L}(Y)^4$ in Eq. \ref{HEREX}, the rest of the solution taking the same form. Similarly, adding more birth events between times $0$ and $T-q_2$ will increase the power of the rightmost $\mathcal{L}(Y)$ term, which again can be dealt with by Laplace transforms. To extend the correlation to all generations such that individuals with age $q_1$ descend from individuals with age $q_2$ (irrespective of generation number) requires summing over all possible generation numbers. This is achieved by replacing the $\mathcal{L}(Y)$ terms with $\sum_{k=0}^\infty\mathcal{L}(Y)^k=(1-\mathcal{L}(Y))^{-1}$. Provided the Laplace convolutions are possible, the pair correlation is still a single integral over variable $r$ and the high dimensionality of the problem can be avoided. 

For more general hereditary structures, the same techniques can be applied, and it is the complexity of the hereditary structure that dictates the degree of integration involved, not the number of generations. Note that the number of different hereditary structures is limited; the internal nodes either preserve or increase the number of propagators through time. Thus any diagram for a paired correlation can only have one or two initiating nodes, for example. This summability and restricted choice of diagram is somewhat analogous to the summability of the `parquet' diagrams observed in \cite{Peliti1986}. The full set of correlation functions up to second order can be found in Appendix C, where the results have also been implemented for the example given in Fig. \ref{QFT_Fig2}.

We note that the path integral for the density function $f_m(\q_m;t)$ in Eq. \ref{DENSPI} differs from the correlation function by the single exponent $i\int \dd p\HS v_T$. Although one can expand this term and perform a similar series expansion to above, this results in a significant increase in the number of terms and it is not clear whether the resulting series can be summed exactly or approximated easily, and finding efficient means of determining the probability density function remains an open problem.

Finally, we note that the action in Eq. \ref{DENSPI} and \ref{MOMPI} are identical, and linear in $v$. This suggests a direct functional integration with respect to $v$ might yield a direct evaluation for $f_m(\q_m;t)$ or $X_m(\q_m;t)$ without recourse to perturbative expansion. Although integration with respect to $v$ can be done, resulting in a constraint upon $u$, which can furthermore be solved, the result requires $m^\textrm{th}$ order functional derivatives which soon become intractable and an exact approach without perturbative expansion is still lacking. An outline of this approach is given in Appendix F.


\section{Binary Fission and Death Processes}
\label{FISSSEC}

We now extend these methods to binary fission, a more general process where two individuals simultaneously arise at the moment the parental individual terminates, a process exemplified by cell division. This process is represented in Fig. \ref{F1_MicroMod}B, where a single cell undergoes a process of cell division and death. Note that at time $t_1$, after three cell divisions, we have two pairs of \emph{twins}, each pair having identical ages. By time $t_2$, one twin has died and one has divided, leaving one pair of twins, and two lone individuals with distinct ages, which we refer to as \emph{singletons}. Fission is thus a degenerate process because a subset of ages are duplicated, and we will show that the ages of singletons and twins need to be tracked separately.

We shall first consider mean-field behavior, then the full probability density for the population-size-age-chart, and finally use path integrals to determine the correlation functions.


\subsection{Kinetic Equations}

We first consider a single species model, before establishing that the age degeneracy is better handled with a two species model. We thus have a death process $A(p) \rightarrow \phi$ occurring at a rate $\mu(p)$ and a fission process of the form $A(p) \rightarrow A(0) + A(0)$ occurring at a rate $\beta(p)$, where $A(p)$ represents a single cell of age $p$. 

Note that age degeneracy is automatically handled in field theoretic formalism. For example, if we apply the number operator $\psi_r^\dag\psi_r$ to the degenerate state $\ket{p,p,q}$, where one age $p$ is duplicated, we obtain the density $\braket{1|\psi_r^\dag\psi_r|p,p,q} = 2\delta(r-p)+\delta(r-q)$ and find the duplicated age is correctly counted.

For the fission process, the operators $\zeta_0$ and $\zeta_d$ are identical to those in Eq. \ref{MICRODEF}, but the birth operator becomes $\zeta_b = \int \dd p \beta(p)[\psi_p^\dag \psi_p-\psi_0^\dag \psi_0^\dag\psi_p]$, where the two $\psi_0^\dag$ operators in the latter term account for the birth of twins, and $\psi_p$ represents termination of the parent.

To study the mean field behavior, we define the mean density $X(p) = \braket{1|\psi_p e^{-\zeta t}|f(0)}$ and differentiate with respect to time, following the methodology used in earlier sections. The derivation is largely the same as that for Eq. \ref{McKendrickInOne}, resulting in a McKendrick-von Foerster-like equation,

\begin{equation} 
\frac{\partial X}{\partial t}+\frac{\partial X}{\partial p} = - \gamma(p)X.
\label{FUNNYMCK}
\end{equation}
Note that the difference between this microscopic model and that encapsulated in Eq. \ref{McKendrickInOne} is that in fission both birth from, and death to, an individual results in their termination. This is reflected in the right-hand side of Eq. \ref{FUNNYMCK}, where the event rate $\gamma(p)=\beta(p)+\mu(p)$ is used. The boundary condition is also modified to account for the duplicated offspring:

\begin{equation}
X(0) = 2\int \dd p \beta(p)X(p).
\label{DegBC}
\end{equation}

Although this formalism works for the mean-field, if we attempt to take the temporal derivative of the pair correlation $X(p,q) = \braket{1|\psi_p\psi_q e^{-\zeta t}|f(0)}$, the following PDE and boundary condition results,

\begin{equation*}
\frac{\partial X}{\partial t}+\frac{\partial X}{\partial p} + \frac{\partial X}{\partial q} = - (\gamma(p)+\gamma(q))X,
\end{equation*}
\begin{equation}
X(0,q) = 2\int \dd r \beta(r)X(q,r) + 2\delta(q)\int \dd r \beta(r)X(0,q,r).
\label{BadBC}
\end{equation}
Note that the boundary condition contains an extra term (compared to Eq. \ref{DegBC}) implicating the third correlation function $X(0,p,r)$ and a delta function $\delta(q)$. These terms arise due to the fact that pairs of individuals can have the same age, and are difficult to deal with analytically. 

To combat this degeneracy, we adopt a multi-species Doi-Peliti paradigm. We treat the two classes (singletons and twins) as separate species, where $A(p)$ represents a singleton cell of age $p$ and $B(p)$ represents a pair of twin cells with identical ages $p$. Then a singleton undergoes a death process $A(p) \rightarrow \phi$ at rate $\mu(p)$ and a fission process $A(p) \rightarrow B(0)$ at rate $\beta(p)$, whereas twins undergo a death process $B(p) \rightarrow A(p)$ at a rate $2\mu(p)$ and a fission process $B(p) \rightarrow A(p) + B(0)$ at a rate $2\beta(p)$. We thus have two pairs of creation and annihilation operators; $\psi_p^\dag,\psi_p$ and $\chi_p^\dag,\chi_p$, respectively. We then represent $m$ singletons by age-chart $\p_m$ and $n$ sets of twins by age-chart $\q_n$, resulting in a general state of the form

\begin{equation}
\ket{\p_m;\q_n} \equiv \psi_{p_1}^\dag\cdots\psi_{p_m}^\dag
\chi_{q_1}^\dag\cdots\chi_{q_n}^\dag\ket{\phi}.
\end{equation}
Here the action of the singleton (resp. twin) operator is the same, irrespective of the current twin (singleton) state. Thus, for example, $\psi_p^\dag\chi_q^\dag\ket{\p_m;\q_n}=\psi_p^\dag\ket{\p_m;q,\q_n} = \ket{p,\p_m;q,\q_n} = \chi_q^\dag\ket{p,\p_m;\q_n} = \chi_q^\dag\psi_p^\dag\ket{\p_m;\q_n}$, and we find the two classes of operators commute with each other (e.g., $[\psi_p^\dag,\chi_q^\dag]=0$). Within each operator class, the usual commutation relations apply; $[\psi_p,\psi_{p'}^\dag]=\delta(p-p')$ and $[\chi_q,\chi_{q'}^\dag]=\delta(q-q')$.

The evolution operators take on the following form:

\begin{eqnarray}
\zeta_0 = & \int \dd p \HS\psi_p^\dag\frac{\partial}{\partial p}\psi_p
+\int \dd p \HS\chi_p^\dag\frac{\partial}{\partial p}\chi_p,
\nonumber\\
\zeta_b = & \int \dd p\HS \beta(p)\left[\psi_p^\dag\psi_p-\chi_0^\dag\psi_p\right]
+2\int \dd p\HS \beta(p) \left[\chi_p^\dag\chi_p-\chi_0^\dag\psi_p^\dag\chi_p\right],
\nonumber\\
\zeta_d = & \int \dd p \HS\mu(p)\left[\psi_p^\dag\psi_p-\psi_p\right]
+2\int \dd p \HS\mu(p)\left[\chi_p^\dag\chi_p-\psi_p^\dag\chi_p\right].
\end{eqnarray}
These operators reflect the microscopic fission process, generalizing the operators in Eq. \ref{MICRODEF} that represent the simpler, non-fission budding birth-death process. For example, the last term $\chi_0^\dag\psi_p^\dag\chi_p$ in $\zeta_b$ represents the event that one individual from a pair of twins divides into a newborn pair of twins; the operator $\chi_p$ represents the annihilation of the pair of twins of age $p$, the term $\chi_0^\dag$ represents the creation of newborn twins of age zero, and the creation operator $\psi_p^\dag$ represents the surviving singleton of age $p$. The coefficient $2$ reflects the fact that either twin of age $p$ can be annihilated. 

We next derive the singleton-twin specific mean field equations given in \cite{Chou2016} as follows. First we define singleton density $A(p) = \braket{1|\psi_p^\dag\psi_p|f(t)}$ and twin density $B(p) =\braket{1|\chi_p^\dag\chi_p|f(t)}$, along with the coherent state

\begin{equation}
\ket{1} \equiv e^{\int dp \psi_p^\dag}e^{\int dq \chi_q^\dag}\ket{\phi}.
\end{equation}
Then differentiating $A(p) = \braket{1|\psi_pe^{-\zeta t}|f(0)}$ and $B(p) = \braket{1|\chi_pe^{-\zeta t}|f(0)}$ with respect to time, using the commutator relations and eigenstate properties of $\ket{1}$, we find

\begin{eqnarray}
&\braket{1|\psi_r\zeta_0|f(t)} = \frac{\partial A}{\partial r},
&\braket{1|\chi_r\zeta_0|f(t)} = \frac{\partial B}{\partial r},\nonumber\\
&\braket{1|\psi_r\zeta_d|f(t)} = \mu(r)(A-2B),\hspace{5mm}
&\braket{1|\chi_r\zeta_d|f(t)} = 2 \mu(r)B,\nonumber\\
&\braket{1|\psi_r\zeta_b|f(t)} = \beta(r)(A-2B),
&\braket{1|\chi_r\zeta_b|f(t)} = 2\beta(r)B-\delta(r)\int \dd p\HS \beta(p)(A+2B),
\end{eqnarray}
which result in the following equations and boundary conditions:

\begin{eqnarray}
&\frac{\partial A}{\partial t}+\frac{\partial A}{\partial r} =
-(\beta(r)+\mu(r))(A-2B),\hspace{5mm}
&\frac{\partial B}{\partial t}+\frac{\partial B}{\partial r} =
-2(\beta(r)+\mu(r))B,\nonumber\\
&A(0) = 0,
&B(0) = \int \dd p \HS (A+2B).
\label{FISSMNFD}
\end{eqnarray}

This approach can also be used to derive higher order correlation functions, which can be found in Appendix D. Note that unlike Eq. \ref{BadBC}, in this multispecies formulation there are no awkward delta functions in the boundary condition (viz. Eq. \ref{FissCorrAppBC}). These mean-field equations agree with the system in \cite{Chou2016} (where solutions can also be found), although unlike \cite{Chou2016}, the derivation above does not require the kinetic equation of the full probability density (Eq. \ref{BULKY}), which we derive next. These two formulations of mean field analysis also serve to show that different Doi-Peliti models can reveal different levels of detail from the same underlying stochastic process.

Now $\frac{1}{m!n!}f_{m,n}(\p_m;\q_n;t)$ is the associated full probability density which has the state representation

\begin{equation}
\ket{f(t)} = \sum_{m,n=0}^\infty\int \frac{\dd \p_m \HS \dd\q_n}{m!n!} \HS f_{m,n}(\p_m;\q_n;t)\ket{\p_m;\q_n}.
\label{TwoTypeGenState}
\end{equation}

We differentiate $f_{m,n}(\p_m;\q_n;t) = \braket{\p_m;\q_n|e^{-\zeta t}|f(0)}$ with respect to time and derive bulk equations in the much same manner as Section \ref{KINSEC} to yield

\begin{eqnarray}
\frac{\partial }{\partial t}f_{m,n} & +\sum_{i=1}^m\frac{\partial}{\partial p_i}f_{m,n} + \sum_{j=1}^n\frac{\partial}{\partial q_j}f_{m,n} = -f_{m,n} \left[\sum_{i=1}^m\left(\beta(p_i)+\mu(p_i)\right) +2\sum_{j=1}^n\left(\beta(q_j)+\mu(q_j)\right)\right]
\nonumber\\
& +\int \dd r \HS \mu(r) f_{m+1,n}(\p_m,r;\q_n;t)
+2\sum_{i=1}^m\mu(p_i)f_{m-1,n+1}(\p_m^{(-i)};\q_n,p_i;t),
\label{BULKY}
\end{eqnarray}

\begin{eqnarray}
f_{m+1,n}(\p_m,0;\q_n;t) & = 0,\nonumber\\
f_{m,n+1}(\p_m;\q_n,0;t) & = \int \dd r \HS \beta(r) f_{m+1,n-1}(\p_m,r;\q_n;t) \nonumber\\
&\hspace{20mm}+ 2\sum_{j=1}^m\beta(p_i)f_{m-1,n+1}(\p_m^{(-i)};\q_n,p_i;t).
\end{eqnarray}
These results agree with the equations obtained via a probabilistic derivation in \cite{Chou2016}.


\subsection{Path Integrals and Correlation Functions}

We next define correlation function $X_{m,n}(\q_m;\bar{\q}_n;t)\dd \q_m\dd \bar{\q}_n$ to be the probability that there are $m$ singletons that can be labeled such that the $i^\textrm{th}$ has age in interval $[q_i,q_i+\dd q_i]$, and $n$ pairs of twins that can be arranged to have ages in intervals $[\bar{q}_i,\bar{q}_i+\dd \bar{q}_i]$. In all that follows, a variable covered by a bar corresponds to a twin property, whereas an unbarred variable refers to singletons. There are two approaches to investigate the correlation functions. Firstly, we can derive kinetic equations for correlation functions in much the same way that Eq. \ref{FISSMNFD}, \ref{FissCorrAppBulk} and \ref{FissCorrAppBC} were derived, which we can then attempt to solve. However, the equations for higher order correlations quickly become complicated (see Appendix D). The second approach develops a suitable path integral representation, which we now consider. To do this we require functional coherent states;

\begin{equation}
\ket{u,\bar{u}} \equiv e^{\int dp \HS u(p)\psi_p^\dag}e^{\int dq \HS \bar{u}(q)\chi_q^\dag}\ket{\phi},
\end{equation}
along with a resolution of the identity;

\begin{equation}
I = \int\mathcal{D}u\mathcal{D}\bar{u} \mathcal{D}v\mathcal{D}\bar{v} \HS e^{-i\int
dq\hspace{0.5mm} (uv+\bar{u}\bar{v})}\ket{iv,i\bar{v}}\bra{u,\bar{u}}.
\label{ResId2}
\end{equation}

Now if we assume that the initial number of singletons and pairs of twins are Poisson distributed with means $\alpha$ and $\bar{\alpha}$, and the ages of these individuals have probability densities $\omega(q)$ and $\bar{\omega(q)}$, respectively, then by using the resolution of the identity between time slices, the correlation function can be written as the following path integral, in much the same way as Eq. \ref{MOMPI} to give the following:

\begin{eqnarray}
X_{m,n}(\q_m;\bar{\q}_n;T) = & \int \mathcal{D}u\mathcal{D}\bar{u} \mathcal{D}v\mathcal{D}\bar{v}
\prod_{j=1}^m iv_T(q_i) \prod_{k=1}^n i\bar{v}_T(\bar{q}_i) \cdot\nonumber\\
&\exp\left\{-i\int \dd q \dd t \left[u\left(\frac{\partial v}{\partial q}+\frac{\partial v}{\partial t}+\gamma (v-2\bar{v})\right)
-\beta\bar{u}^0v \right]\right\}\cdot\nonumber\\
& \exp\left\{-i\int \dd q \dd t \left[\bar{u}\left(\frac{\partial \bar{v}}{\partial q}+\frac{\partial \bar{v}}{\partial t}+2\gamma \bar{v}\right)
-2\beta(u+1)\bar{u}^0\bar{v} \right]\right\}\cdot\nonumber\\
& \exp\left\{ -i\int \dd q \hspace{1mm}(u_0v_0+\bar{u}_0\bar{v}_0) + \alpha\int \dd q \HS \omega u_0 + \bar{\alpha}\int \dd q\HS \bar{\omega}\bar{u}_0 \right\}.
\label{MOMFISSPI}
\end{eqnarray}

\begin{figure}[t!]
\begin{center}
\includegraphics[width = 14cm,right]{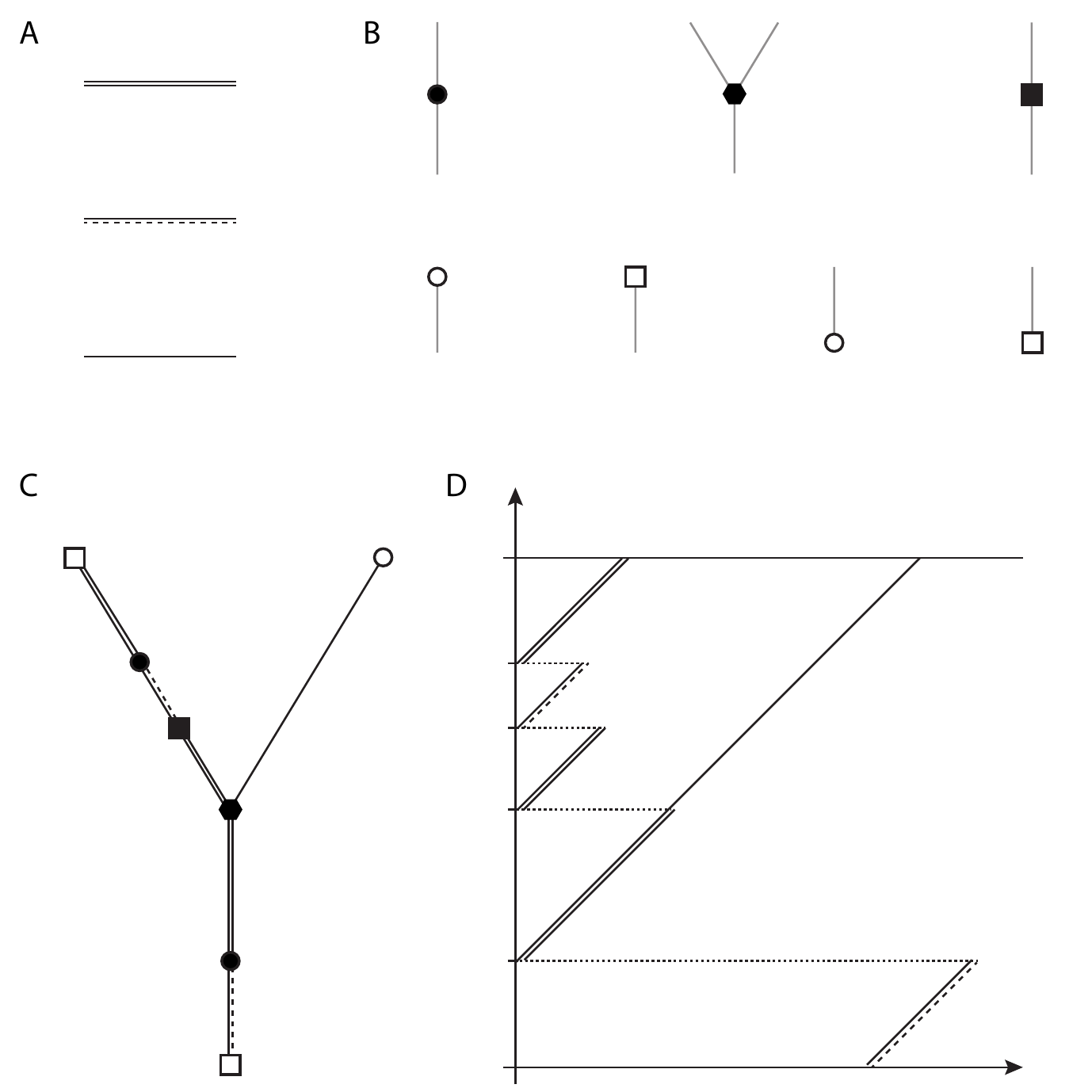}
\put(-37,-4){Age}
\put(-204,214){Time}
\put(-177,199){$\bar{q}_2$}
\put(-67,199){$q_1$}
\put(-222,192){$T$}
\put(-242,155){$T-\bar{q}_2$}
\put(-223,132){$\tau_2$}
\put(-223,102){$\tau_1$}
\put(-242,46){$T-q_1$}
\put(-83,0){$r$}
\put(-217,0){$0$}
\put(-358,355){$G_{uv} \sim \hat{S}$}
\put(-372,305){$G_{\bar{u}v} \sim 2\hat{S}(1-\hat{S})$}
\put(-360,255){$G_{\bar{u}\bar{v}} \sim \hat{S}^2$}
\put(-237,369){$\bar{u}^0$}
\put(-237,353){$v$}
\put(-263,361){$\beta(q)$}
\put(-135,371){$\bar{u}^0$}
\put(-122,371){$u$}
\put(-127,353){$\bar{v}$}
\put(-159,361){$2\beta(q)$}
\put(-19,370){$\bar{u}^0$}
\put(-19,352){$\bar{v}$}
\put(-52,361){$2\beta(q)$}
\put(-237,288){$v$}
\put(-165,284){$\bar{v}$}
\put(-92,277){$u$}
\put(-106,260){$\alpha\omega(q)$}
\put(-20,279){$\bar{u}$}
\put(-32,260){$\bar{\alpha}\bar{\omega}(q)$}
\caption{Feynman diagram and associated Age-Time diagram for fission process. (A) Three classes of propagator. (B) Three classes of internal nodes, two initiating and two terminating nodes. (C) Feynman diagram consisting of one initiating node, two terminating nodes, three degree two nodes and one degree three node. (D) The associated Age-Time diagram. Individuals of age $\bar{q}_2$ are twins with generation number $4$, descending from generation $1$, which contain singletons of age $q_1$.}
\label{FissFeyn}
\end{center}
\end{figure}
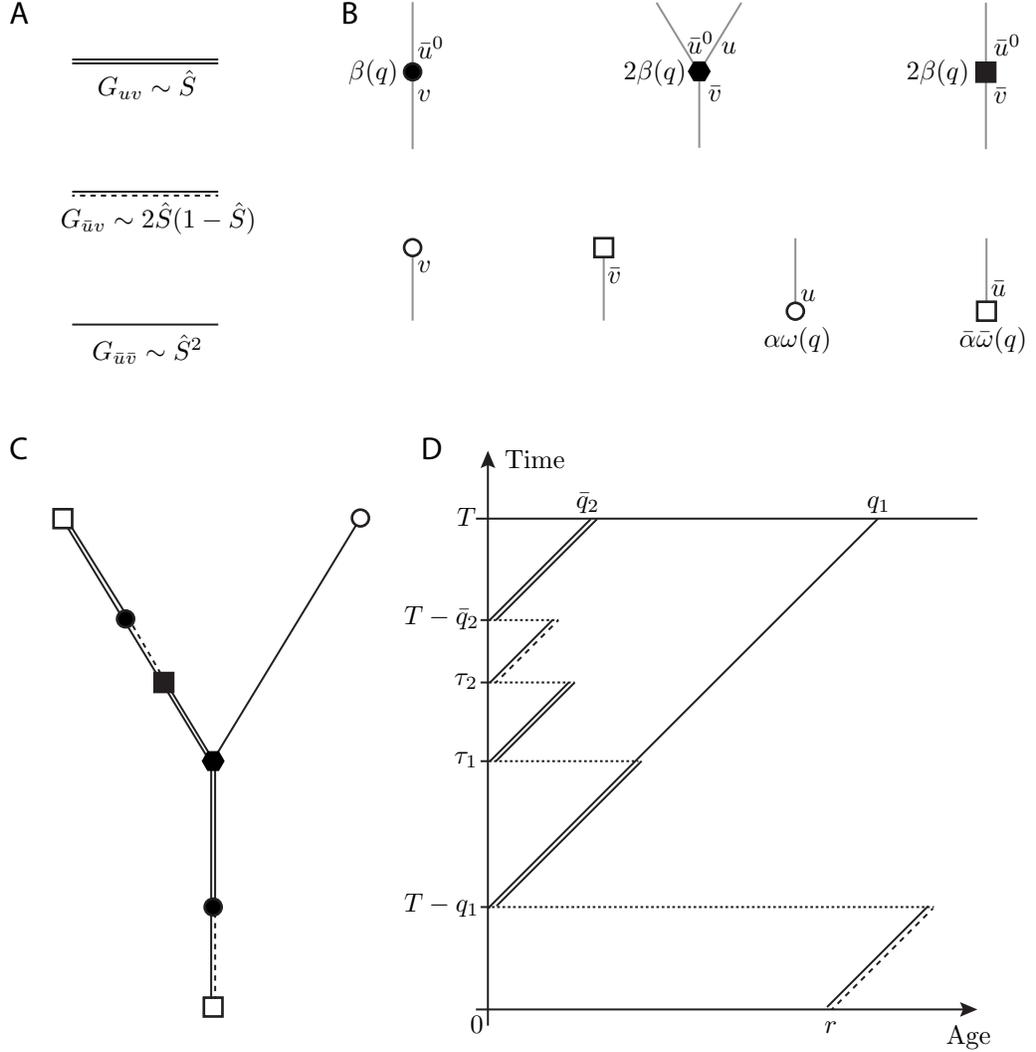

Now, as before, we expand the birth terms $\beta\bar{u}^0v$ and $2\beta(u+1)\bar{u}^0\bar{v}$, along with the initial distribution terms $\bar{\alpha}\int \dd q\HS \omega u_0$ and $\bar{\alpha}\int \dd q\HS \bar{\omega}\bar{u}_0$. We again split the (latter) birth term giving three types of birth events, meaning we have three types of internal nodes in the associated Feynman diagram, as indicated in Fig. \ref{FissFeyn}B. The two initial distribution terms mean we have two classes of initiating nodes, and the singleton and twin class also mean there are two types of terminating node. We shall construct propagators in terms of functional derivatives of the following generating functional:

\begin{eqnarray}
Z(J,K,\bar{J},\bar{K}) = & \int \mathcal{D}u\mathcal{D}\bar{u} \mathcal{D}v\mathcal{D}\bar{v}
\exp\left\{-i\int \dd q \dd t \left[u\left(\frac{\partial v}{\partial q}+\frac{\partial v}{\partial t}+\gamma (v-2\bar{v})+J\right)
 \right]\right\}\cdot\nonumber\\
& \exp\left\{-i\int \dd q \dd t \left[\bar{u}\left(\frac{\partial \bar{v}}{\partial q}+\frac{\partial \bar{v}}{\partial t}+2\gamma \bar{v}+\bar{J}\right)
\right]\right\}\cdot\nonumber\\
&\exp\left\{ -i\int \dd q \hspace{1mm}(u_0v_0+\bar{u}_0\bar{v}_0)  +i\int \dd q \dd t \HS (vK+\bar{v}\bar{K})\right\}.
\end{eqnarray}

Functionally integrating over $u$ and $\bar{u}$ results in the conditions $\frac{\partial v}{\partial q}+\frac{\partial v}{\partial t}+\gamma (v-2\bar{v})+J=0$ and $\frac{\partial \bar{v}}{\partial q}+\frac{\partial \bar{v}}{\partial t}+2\gamma \bar{v}+\bar{J}=0$. Furthermore, integration over $u_0$ and $\bar{u}_0$ results in boundary conditions $v_0=0$ and $\bar{v}_0=0$, resulting in solutions:

\begin{eqnarray}
\bar{v}(q,t) = -\int_0^{t}\bar{J}(q-t+s,s)e^{-\int_s^t2\gamma(q-t+x)dx}\dd s,\nonumber\\
v(q,t) = -\int_0^{t}\left[J(q-t+s,s)-2\bar{v}(q-t+s,t)\gamma(q-t+s)\right]
e^{-\int_s^t\gamma(q-t+x)dx}\dd s,
\end{eqnarray}
with generating functional
\begin{equation}
Z(J,K,\bar{J},\bar{K}) = \exp\left\{i\int \dd q \dd t \HS (vK+\bar{v}\bar{K})\right\}.
\end{equation}

We then find that the four possible propagators are:

\begin{eqnarray}
\fl
G_{uv}(p,\tau;p',\tau') = \left.\frac{i\delta^2Z(J,K,\bar{J},\bar{K})}{\delta J(p;\tau)\delta K(p';\tau')} \right|_{J,\bar{J},K,\bar{K}\equiv 0}=\delta((p'-\tau')-(p-\tau))\hat{S}(p,p')\theta(\tau-t),\nonumber\\
\fl
G_{u\bar{v}}(p,\tau;p',\tau') = \left.\frac{i\delta^2Z(J,K,\bar{J},\bar{K})}{\delta J(p;\tau)\delta \bar{K}(p';\tau')} \right|_{J,\bar{J},K,\bar{K}\equiv 0}=0,\nonumber\\
\fl
G_{\bar{u}v}(p,\tau;p',\tau') = \left.\frac{i\delta^2Z(J,K,\bar{J},\bar{K})}{\delta \bar{J}(p;\tau)\delta K(p';\tau')} \right|_{J,\bar{J},K,\bar{K}\equiv 0}=\delta((p'-\tau')-(p-\tau))2\hat{S}(p,p')(1-\hat{S}(p,p'))\theta(\tau-t),\nonumber\\
\fl
G_{\bar{u}\bar{v}}(p,\tau;p',\tau') = \left.\frac{i\delta^2Z(J,K,\bar{J},\bar{K})}{\delta \bar{J}(p;\tau)\delta \bar{K}(p';\tau')} \right|_{J,\bar{J},K,\bar{K}\equiv 0}=\delta((p'-\tau')-(p-\tau))\hat{S}(p,p')^2\theta(\tau-t),
\end{eqnarray}
where we have used the survival function $\hat{S}(p,p')=\exp\left\{-\int_p^{p'}\gamma(x)\dd x\right\}$. Note the distinction from the survival function $S(p,p') = \exp\left\{-\int_p^{p'} \mu(x)\dd x\right\}$ employed in previous sections. This is to be expected; in the simple mode of birth, where each parent can have multiple birth events, survival is just predicated on death not occurring over the interval $[p,p']$, encapsulated by the death rate $\mu(x)$. In the fission process, however, birth induces death of the parent, so survival requires that both birth and death does not occur across the interval $[p,p']$ prior to the birth event at time $t'$, hence the event rate $\gamma(x)= \beta(x)+\mu(x)$ is employed. The propagator function $\hat{S}^2$ thus represents the probability that a pair of twins do not die or give birth across an interval. The function $2\hat{S}(1-\hat{S})$ represents the Binomial probability that one of a pair of twins dies or gives birth and the remaining singleton survives.

Thus we have three non-trivial propagators, resulting in three types of edges in the associated Feynman diagram, as highlighted in Fig. \ref{FissFeyn}A (see also Appendix E).

Consider then the Feynman diagram of Fig. \ref{FissFeyn}C, which is directly mapped onto the Age-Time chart in Fig. \ref{FissFeyn}D. We have a founder generation (generation number $0$) of twins of age $r$, giving rise to generation number $1$ twins at time $T-q_1$. At time $\tau_1$ one of the twins gives birth and dies, leaving the singleton of age $q_1$ at time $T$. We have two more generations giving birth; one a twin giving birth at time $\tau_2$ and one a singleton at time $T-\bar{q}_2$, resulting in twins of age $\bar{q}_2$ at time $T$. The integral corresponding to this diagram thus contributes to the paired correlation function $X_{1,1}(q_1;\bar{q}_2;t)$ with a term of the form:

\begin{eqnarray}
\hat{X}_{1,1}(q_1;\bar{q}_2;T) & = \bar{\alpha}  \hat{S}(\bar{q}_2)^2\hat{S}(q_1)\int_0^\infty \dd r \HS\bar{\omega}(r)\hat{S}_2(r,r+[T-q_1])\beta(r+[T-q_1])\cdot\nonumber\\
& \hspace{10mm}\int\displaylimits_{T-q_1<\tau_1<\tau_2<T-\bar{q}_2} \hspace{-10mm}\dd \tau_1 \dd \tau_2 Y_1(\tau_1-[T-q_1]) Y_2(\tau_2-\tau_1) Y_3([T-\bar{q}_2]-\tau_2)\nonumber\\
& = \bar{\alpha}\hat{S}(\bar{q}_2)^2\hat{S}(q_1)\mathcal{L}^{-1}_{q_1-\bar{q}_2} (\mathcal{L}(Y_1) \mathcal{L}(Y_2) \mathcal{L}(Y_3))\cdot\nonumber\\
& \hspace{22mm}\int_0^\infty \dd r \HS\hat{\omega}(r)\hat{S}_2(r,r+[T-q_1])\beta(r+[T-q_1]),
\end{eqnarray}
where we have used the shorthand notation $\hat{S}_2=2\hat{S}(1-\hat{S})$, $Y_1 = 2\beta\hat{S}$, $Y_2 = 2\beta\hat{S}^2$ and $Y_3=2\beta\hat{S}(1-\hat{S})$. We have also used Laplace transforms to calculate the $\tau_1,\tau_2$ convolution integral. 

Note that we can place as many generations as we like between times $T-q_1$ and $T-\bar{q}_2$. These will give rise to a multiple convolution of terms where $\mathcal{L}(Y_2)\mathcal{L}(Y_3)$ is replaced with powers of $\mathcal{L}(Y_2)$ and $\mathcal{L}(Y_3)$. If we sum over all possibilities, we just replace $\mathcal{L}^{-1}_t (\mathcal{L}(Y_1) \mathcal{L}(Y_2) \mathcal{L}(Y_3))$ with $\mathcal{L}^{-1}_t (\mathcal{L}(Y_1) (1-(\mathcal{L}(Y_2)+\mathcal{L}(Y_3)))^{-1})$ and find that the correlation functions can be entirely summed. Furthermore, just like the simple budding mode of birth, we find that the internal nodes of Feyman diagrams either preserve or increase the number of propagators through time, meaning that the number of initiating nodes is limited by the degrees of the correlation function, limiting the number of possible hereditary structures that need to be considered. 


\section{Conclusions}

In this work we have used age-structured population modeling to highlight Doi-Peliti methods of population inference. In particular, we have generalized the formulation of coherent states used in \cite{Doi1976, Doi1976b} to construct a path integral formulation similar to that of \cite{Peliti1985} without recourse to lattice methods. This technique was applied to age structured systems, firstly providing an efficient means of deriving kinetic equations for features of interest, and secondly demonstrated that perturbative expansions of path integrals can be calculated exactly, providing a means to construct age-structured correlation functions. These expansion methods do not so easily sum for probability density functions, which generally need to be approximated by truncation methods \cite{Raghib2011,Rogers2011}; finding efficient means of solving hierarchical equations such as Eq. \ref{RHO0} remains an open problem. 

Although summing the perturbation series is possible, it involves calculating terms such as $\mathcal{L}_{q_2-q_1}^{-1}(\mathcal{L}(\beta)(1-\mathcal{L}(Y))^{-1})$, which will appear in Eq. \ref{HEREX} once it is modified to sum over all generations. The Laplace inversion here will likely require a Bromwich integral and the resulting expressions can be complicated (an example of this kind of calculation can be found in \cite{Chou2016}). However, we see from Fig. \ref{QFT_Fig2}B that for any given time the number of generations contributing to correlation functions are limited, and summing over a subset of generations can prove easier and just as effective. 

The path integral construction presented assumes that the initial population size is Poisson distributed. This naturally fits with the coherent state framework that was used and is the most efficient formulation. However, the machinery can be adapted to enable more general initial conditions, although some of the efficiency will be lost, resulting in integrals over a range of initial conditions.

The age structured systems we have introduced can be considered as spatial systems where age is a convective term increasing with time. Other spatial properties, such as the molecular diffusive phenotype commonly analyzed in reaction-diffusion systems, can readily be incorporated into the models we have considered by adding suitable diffusive terms into the evolution equations, although this was not explored in this work. 

There is a need to develop the Doi-Peliti approach presented above to incorporate non-linear effects. For example, more realistic patterns of growth often involve population-size constraints such as a carrying capacity. This was noticed early in the development of population dynamics by Verhulst who modeled the growth deterministically with logistic functions \cite{Verhulst1838}. In \cite{Greenman2016, Chou2016} population-size (and age-dependent) birth and death rates, $\beta_n(q)$ and $\mu_n(q)$, were analyzed for a fully stochastic model, although the corresponding kinetic equations are significantly complicated by population-size dependency. For example, the mean-field equations are only equivalent to the McKendrick-von Foerster equation when the rates are independent of population size. 

Another non-linear effect of interest arises when fission processes are mediated by cell size rather than age \cite{Webb2008}, for which the non-linearity can manifest as chaos \cite{Brzezniak2014}. The second quantization techniques we have introduced need to be generalized to cater for such effects. The linear action of operators in field theory suggest that dealing with non-linear effects will be difficult, although the success of approaches in the area of quantum chaos give some cause for optimism \cite{Casati2006}. The models we have introduced model interactions at the level of a single cell; population-size effects could be modeled by requiring that all cells interact with each other, for example. It seems unlikely that corresponding path integrals will be either simple or exact, and perturbative methods of approximation or semi-classical techniques will likely be required \cite{Assaf2010}. 

Exploring such avenues will be required if we are to further develop these methods in more realistic situations. However, the machinery we have introduced provides an initial framework that can efficiently deal with many age-structured models, offering an approach that avoids some of the complications presented when more classical probabilistic techniques are employed \cite{Greenman2016, Chou2016}. Furthermore, path integral formulations of this machinery enabled explicit relationships between correlation functions and inheritance structure to be identified, results that are not obvious when applying more probabilistic approaches.


\appendix
\addcontentsline{toc}{section}{Appendix A: Resolution of the Identity}
\setcounter{section}{1}
\section*{Appendix A: Resolution of the Identity}

We outline why Eq. \ref{ResId} is equivalent to the identity operator. More information about using coherent states and resolutions of the identity can be found in \cite{Fradkin1991}. The path integral in Eq. \ref{ResId} is understood in the following sense:

\begin{equation}
I = \lim_{\scriptsize{\begin{array}{c}P \rightarrow \infty \\ \epsilon \rightarrow 0 \end{array}}}
\prod_{i=-P/\epsilon}^{P/\epsilon}\int \frac{\epsilon}{2\pi}\dd [u(p_i)]\HS\dd [v(p_i)]e^{-i\epsilon \sum_{i=-P/\epsilon}^{P/\epsilon}u(p_i)v(p_i)}\ket{iv}\bra{u},
\label{PAD}
\end{equation}
where age has been discretized over the interval $[-P,P]$ such that $p_i=i\epsilon$.

To interpret the coherent states $\bra{u}$ and $\ket{iv}$ we need to use a discretized form of fundamental states $\ket{\p_n}$. To do this we use an occupation number state representation. For example, suppose we have a coarse grain resolution of age so that there are five possible values, $p_{-2}$, $p_{-1}$, $p_0$, $p_1$ and $p_2$. Suppose, furthermore, that we have a discretized state $\ket{\p_4} =\ket{p_0,p_{-2},p_0,p_1}$ comprising the ages of four
individuals. Then we write $\ket{\p_4} \equiv \ket{1,0,2,1,0}_0$, where the subscript $0$ indicates an occupation number state representation. Two of the four individuals have middle age $p_0$, and so the middle occupation number is $2$ in the occupation state. Then we expand and discretize a general coherent state $\ket{g}$ as:

\begin{eqnarray}
\ket{g} & = e^{\int dp \HS g(p)\psi_p}\ket{\phi}=e^{\epsilon\sum_i g(p_i)\psi_{p_i}}\ket{\phi} =\sum_{n=0}^\infty
\frac{\epsilon^n}{n!}\left[ \sum_{i=-P/\epsilon}^{P/\epsilon}g(p_i)\psi_{p_i} \right]^n\ket{\phi}
\nonumber\\
&=\sum_{n=0}^\infty \frac{\epsilon^n}{n!}\sum_{\left\{{\x: \sum_i x_i=n}\right\}}{n \choose \x}\prod_{i=-P/\epsilon}^{P/\epsilon}g(p_i)^{x_i}\ket{\x}_0,
\label{ROI}
\end{eqnarray}
where we use the notation ${n \choose \x}\equiv
\frac{n!}{x_{-P/\epsilon}!\cdots x_{P/\epsilon}!}$. Then if we
similarly define $\x!\equiv x_{-P/\epsilon}!\cdots x_{P/\epsilon}!$ and use Eq. \ref{ROI} to obtain a discretized version of $\ket{iv}\bra{u}$, Eq. \ref{PAD} can be written as

\begin{eqnarray}
I = \lim_{\scriptsize{\begin{array}{c}P \rightarrow \infty\\ \epsilon \rightarrow 0\end{array}}} &
\sum_{m,n=0}^\infty\epsilon^{m+n}\sum_{\left\{\scriptsize{\begin{array}{c}\x,\y: \\ \sum_i x_i=n\\ \sum_j y_j=m\end{array}}\right\}}\frac{1}{\x!\y!}
\cdot\nonumber\\
& \prod_{i=-P/\epsilon}^{P/\epsilon}
\left[\int \frac{\epsilon}{2\pi}\dd [u(p_i)]\HS\dd [v(p_i)]e^{-i\epsilon u(p_i)v(p_i)}u(p_i)^{x_i}(iv(p))^{y_i}\right]\ket{\y}_0\bra{\x}_0.
\label{IEQ}
\end{eqnarray}

Now, integration by parts establishes the following identity \cite{Peliti1985,Gardiner1985}:

\begin{equation}
\int \frac{\dd u \HS \dd v}{2\pi}e^{-iuv}u^m(iv)^n 
=\int \dd u \HS u^m \left(-\frac{\partial}{\partial u}\right)^n\delta(u)
= m!\HS \delta_{mn},
\label{KronInt}
\end{equation}
which can be used to simplify Eq. \ref{IEQ}, giving

\begin{equation}
I = \lim_{\scriptsize{\begin{array}{c}P \rightarrow \infty\\ \epsilon \rightarrow 0\end{array}}}
\sum_{m=0}^\infty \epsilon^m\sum_{\left\{{\x: \\ \sum_i x_i=m}\right\}}\frac{1}{\x!}\ket{\x}_0\bra{\x}_0.
\end{equation}
Each occupancy vector $\ket{\x}_0$ with total occupation number $m$ corresponds to ${m \choose \x}$ possible states $\ket{\p_m}$. After taking the continuum limit we find
\begin{equation}
I = \sum_{m=0}^\infty\int \frac{\dd \p_m}{m!}\ket{\p_m}\bra{\p_m}.
\end{equation}
Thus we have recovered Eq. \ref{RESID}; a standard resolution of the identity, as required. 


\setcounter{section}{2}
\section*{Appendix B: Alternative Path Integral Formulation}
\addcontentsline{toc}{section}{Appendix B: Alternative Path Integral Formulation}
\setcounter{equation}{0}

The fundamental property that enables the path integral formulation given in Eq. \ref{MatPI} is the resolution of the identity described in Appendix A. This relies on the fundamental integral representation of the Kronecker delta function given in Eq. \ref{KronInt}, which was also the formulation used by Peliti \cite{Peliti1985}. However, there also exists the following expression, which can be established by converting to polar coordinates and integrating the subsequent gamma distribution:

\begin{equation}
\int \frac{\dd z \HS \dd z^*}{2\pi i}e^{-zz^*}z^mz^{*n} = m!\delta_{mn}.
\end{equation}
In a manner largely identical to that in Appendix A, this can be used to construct a resolution of the identity more commonly seen in quantum mechanics:

\begin{equation}
I = \int\mathcal{D}f\mathcal{D}f^*e^{-\int dp f^*(p)f(p)}\ket{f}\bra{f}.
\end{equation}

In much the same way as the derivation of Eq. \ref{OperEx3}, we can use eigenstate properties to determine the action of operators $\zeta_0$, $\zeta_b$ and $\zeta_d$ upon coherent states $\ket{f}$ and $\ket{g}$, where $f$ and $g$ are possibly complex functions, to give the following:

\begin{eqnarray}
\braket{f|\zeta_0|g} & = \braket{f|g}\int \dd p \HS f^*(p)\frac{\partial g}{\partial p}(p),\nonumber\\
\braket{f|\zeta_b|g} & = \braket{f|g}(1-f^*(0))\int \dd p\HS \beta(p)f^*(p)g(p),\nonumber\\
\braket{f|\zeta_d|g} & = \braket{f|g}\int \dd p \HS \mu(p)g(p)(f^*(p)-1).
\end{eqnarray}
We then use these to construct the incremental term

\begin{equation}
\braket{f|e^{-\zeta \epsilon}|g} = \exp\left\{-\epsilon \int dp \left[ -\frac{f^*g}{\epsilon} + f^*\frac{\partial g}{\partial p}+ (1-f^*(0))\beta f^*g +\mu g(f^*-1)\right]\right\},
\end{equation}
and thus arrive at the following path integral formulation:

\begin{eqnarray}
\braket{f_T|e^{-\zeta T}|f_0} = \int
\mathcal{D}f\mathcal{D}f^* \exp & \left\{-\int \dd p \dd t\left [f^*(f_t+f_p)+\beta(1-f^*(0,t))ff^*+\mu f(f^*-1)\right]\right\}\cdot\nonumber\\
& \exp\left\{\int \dd pf^*(p,T)f(p,T)\right\}.
\end{eqnarray}
This provides an alternative path integral formulation to Eq. \ref{MatPI}, although we did not explore this representation any further.


\setcounter{section}{3}
\section*{Appendix C: Correlation Functions up to Second Order}
\addcontentsline{toc}{section}{Appendix C: Correlation Functions up to Second Order}
\setcounter{equation}{0}

Here we use the path integral expansion to construct correlation functions up to order $2$ for the process described in Section \ref{Numerics}.

We let $g(t,k,\theta) = \theta^{-k}\Gamma (k)^{-1}t^{k-1} e^{- t/\theta}$ denote the gamma probability density function and $G(t,k,\theta)$ the corresponding cumulative function. We assume death rate $\mu(q) = \frac{g(q,k,\theta)}{1-G(q,k,\theta)}$ and birth rate $\beta(q) = c q^z\mu(q)$. For the implementation in Fig. \ref{QFT_Fig2} we used values $c = 1.2$, $z=0.2$, $k = 16$ and $\theta = \frac{1}{4}$. We also had Poisson distributed initial population size with mean $5$, and initial age distribution $\omega(q) = g(q,4,\frac{1}{4})$. 

We let $I(t,a,b) = \int_0^tx^{a-1}(1-x)^{b-1}\dd x$ denote the incomplete beta function, we let $Y(t) = \beta(t) S(0,t)$, and define $C=\frac{c\Gamma(z+k)}{\theta^z\Gamma(k)}$. Then we  obtain the following convolution:

\begin{equation}
\fl
R_n(r,t) = \mathcal{L}_t^{-1}(\mathcal{L}(Y(t+r))\mathcal{L}(Y(t))^{n-1})
=\left\{ 
\begin{array}{l l}
C^ng(t+r,(z+k)n,\theta)I(\frac{t}{t+r},(z+k)(n-1),z+k), & n > 1,\\
Y(t+r), & n = 1.
\end{array}
\right.
\end{equation}
Note also that $R_0(0,t)=\delta(t)$.

We let $X^{(\ell)}_1(q,t)$ denote the mean density of the $\ell^\textrm{th}$ generation individuals (as plotted in Fig. \ref{QFT_Fig2}B). Then we find that:

\begin{equation}
X^{(\ell)}_1(q,T) =\left\{ 
\begin{array}{l l}
\alpha\omega(q-T)S(q-T,q), & \ell = 0,\\
\alpha\int_0^\infty\dd r\HS \frac{\omega(r)}{S(0,r)}R_\ell(r,T-q), & \ell \ge 1,
\end{array}
\right.
\end{equation}

Now if we let $X^{(\ell,m)}_2(q_1,q_2;T)$ denote the pair correlation function such that ages $q_1$ and $q_2$ correspond to individuals in generations $\ell$ and $m$, respectively, such that both descend from distinct founder individuals (i.e. generation $0$), then the Feynman diagram consists of two disconnected components and we find the simple relationship

\begin{equation}
X^{(\ell,m)}_2(q_1,q_2;T)=X^{(\ell)}_1(q_1;T)X^{(m)}_1(q_2;T).
\end{equation}

Finally, we let $X^{(n,\ell,m)}_2(q_1,q_2;T)$ denote the pair correlation such that $q_1$ and $q_2$ denote the ages of a pair of individuals that descend from the same individual in generation $n$, such that the two correlated individuals belong to generations $n+\ell$ and $n+m$, respectively. Then we have the following cases.

For $n,\ell,m \ge 1$ the two individuals lie on distinct branches of a hereditary tree, neither being a founder individual, and we find that:

\begin{eqnarray}
X^{(n,\ell,m)}_2(q_1,q_2;T) = & \alpha S(q_1) S(q_2) \int\displaylimits_0^\infty \dd r \HS\frac{\omega(r)}{S(r)} \hspace{-5mm}\int\displaylimits_0^{\min(T-q_1,T-q_2)}\hspace{-5mm}\dd t \HS R_n(r,t) \hspace{-2mm}\int\displaylimits_0^{[T-q_1]-t} \hspace{-4mm} \dd u \int\displaylimits_0^{[T-q_2]-t} \hspace{-4mm} \dd v\cdot\nonumber\\
& \frac{Y([T-q_1]-t-u)Y([T-q_2]-t-v)}{S(\min([T-q_1]-t-u,[T-q_2]-t-v))}R_{\ell-1}(0,u)R_{m-1}(0,v).
\end{eqnarray}

For the case $X^{(n,0,m)}_2(q_1,q_2;T)$ we find that the generation $n+m$ individual with age $q_2$ is a direct descendant of the generation $n$ individual with age $q_1(>q_2)$, where we obtain the following (the case for $X^{(n,m,0)}_2(q_1,q_2;T)$ is symmetrically similar):

\begin{eqnarray}
X^{(n,0,m)}_2(q_1,q_2;T) = \alpha S(q_1) & S(q_2)\mathcal{L}^{-1}_{q_1-q_2}(\mathcal{L}(\beta)\mathcal{L}(R_{m-1}(0,t))) \int\displaylimits_0^\infty \dd r \HS\frac{\omega(r)}{S(r)} R_n(r,T-q_1).
\end{eqnarray}

For the case $X^{(0,\ell,m)}_2(q_1,q_2;T)$ that the individuals with ages $q_1,q_2$ descend from a founder individual (generation $0$), we find that:

\begin{eqnarray}
X^{(0,\ell,m)}_2(q_1,q_2;T) = \alpha S(q_1) & S(q_2) \int\displaylimits_0^\infty \dd r \HS\frac{\omega(r)}{S(r)^2}\int\displaylimits_0^{T-q_1}\dd u \int\displaylimits_0^{T-q_2} \dd v \HS\frac{Y(r+u)Y(r+v)}{S(\min(r+u,r+v))} \cdot\nonumber\\
& R_{\ell-1}(0,[T-q_1]-u)R_{m-1}(0,[T-q_2]-v).
\end{eqnarray}

Finally, for the case $X^{(0,0,m)}_2(q_1,q_2;T)$ that $q_1$ is the age of a founder individual, and $q_2$ is the age of a generation $m$ individual (so $q_1>T>q_2$), we get the following (the case of $X^{(0,\ell,0)}_2(q_1,q_2;T)$ is symmetrically similar):

\begin{eqnarray}
X^{(0,0,m)}_2(q_1,q_2;T) = \alpha S(q_1) & S(q_2) \frac{\omega(q_1-T)}{S(q_1-T)}\mathcal{L}^{-1}_{T-q_2}(\mathcal{L}(\beta(q_1-T+t))\mathcal{L}(R_{m-1}(0,t))).
\end{eqnarray}

The example described in more detail in Eq. \ref{HEREX} is precisely $X^{(3,3,0)}_2(q_1,q_2;T)$. To obtain the analytic variance given in Fig. \ref{QFT_Fig2}Cii, the expressions above were summed for all generations up to $5$ and substituted into Eq. \ref{VarEq}.


\setcounter{section}{4}
\section*{Appendix D: Correlation Function Equations for Fission}
\label{AppFissEq}
\addcontentsline{toc}{section}{Appendix D: Correlation Function Equations for Fission}
\setcounter{equation}{0}
\label{FissCorrApp}

We have the following equation for the correlation function obtained by differentiating $X_{m,n}(\p_m;\q_n;t)=\braket{1|\psi_{p_1}\dots\psi_{p_m} \chi_{q_1}\dots \chi_{q_n}e^{-\zeta t} |f(0)}$ with respect to time and using commutation relations in much the same way as the derivation of Eq. \ref{FISSMNFD}:

\begin{eqnarray}
\frac{\partial X_{m,n}}{\partial t} + \sum_{i=1}^m\frac{\partial X_{m,n}}{\partial p_i}  + \sum_{j=1}^n\frac{\partial X_{m,n}}{\partial q_j}
+\left[\sum_{i=1}^m\gamma(p_i)+2\sum_{j=1}^n\gamma(q_j)\right] X_{m,n} & = &\nonumber\\
&&\hspace{-40mm}2\sum_{i=1}^m\gamma(p_i)X_{m-1,n+1}(\p_m^{(-i)};\q_n,p_i;t),
\label{FissCorrAppBulk}
\end{eqnarray}
with boundary condition
\begin{eqnarray}
X_{m,n}(\p_m;\q_{n-1},0;t) & = & \int\dd r \HS \beta (r) X_{m+1,n-1}(\p_m,r;\q_{n-1};t) + 2\int \dd r \HS \beta (r) X_{m,n}(\p_m;\q_{n-1},r;t)
\nonumber\\
&& +2\sum_{i=1}^m\beta (p_i) X_{m-1,n}(\p_m^{(-i)};\q_{n-1},p_i;t).
\label{FissCorrAppBC}
\end{eqnarray}

\setcounter{section}{5}
\section*{Appendix E: Feynman Diagram Summary}
\addcontentsline{toc}{section}{Appendix E: Feynman Diagram Summary}
\label{FEYNAPP}
\setcounter{equation}{0}

\subsection*{Budding Birth}

The construction of Feynman diagrams and corresponding Age-Time diagrams is required to represent terms arising in the expansion of $X_m(\q_m;t)$ (see also Fig. \ref{Hered_Exam}), each term begin an integral over a product of \emph{factors}. These are outlined below:

\begin{itemize}
  \item There are a number $I\le m$ of initiating nodes, each associated with \emph{factor} $\alpha\omega(p_i)$, $i=1,...,I$, where $p_i$ represents the age of a founder individual.
  \item There are a number $J$ of internal nodes, of degrees $2$ or $3$. Collectively, they are associated with times $0 \le t_1 \le t_2 \le\dots \le t_J \le T$.
  \item There are $m$ terminating nodes, all associated with time $T$. The $\ell^\textrm{th}$ is also associated with age $q_\ell$, $\ell=1,2,\dots,m$. We take the first node connected below the $\ell^\textrm{th}$ terminating node that does \emph{not} involve any edge to the right of a node of degree $3$. 
  \begin{itemize}
  \item If it is the $j^\textrm{th}$ internal node, we associate \emph{factor} $\delta(q_\ell - (T-t_j))$. 
  \item If it is the $i^\textrm{th}$ initiating node, we associate \emph{factor} $\delta(q_\ell - (T+p_i))$.
  \end{itemize}
  \item The Age-Time diagram is constructed as follows. Each edge connecting a pair of nodes in the Feynman diagram is associated with initial and final age-time pairings $(p,t)$ and $(p',t')$, respectively. There is a corresponding line from $(p,t)$ to $(p',t')$ in the Age-Time diagram (Fig. \ref{Hered_Exam}C). The pairings can be calculated by working from the initiating nodes, upwards, with the following observations.
  \begin{itemize}
  \item Edges connected to the $i^\textrm{th}$ initial node are given an initial age-time pairing of $(p_i,0)$.
  \item If the starting node of an edge is the $j^\textrm{th}$ internal node, which also has degree $2$, the initial age-time pairing is $(0,t_j)$.
  \item If the starting node of an edge is the $j^\textrm{th}$ internal node, which also has degree $3$, and the edge extends up and to the left from the node, the initial age-time pairing is $(0,t_j)$ (representing birth of a new individual).
  \item If the starting node of an edge is the $j^\textrm{th}$ internal node, which also has degree $3$, and the edge extends up and to the right from the node, the initial age-time pairing is $(p'',t_j)$, where $p''$ is the final age of the edge below the node (representing the age of the parent who gave birth to a new individual).
  \item The final age-time pairing for an edge where the final node is the $j^\textrm{th}$ internal node is $(p-t+t_j,t_j)$, where $(p,t)$ is the initial age-time pairing. 
  \item Edges connected to the $\ell^\textrm{th}$ terminating node are given a final age-time pairing $(q_\ell,T)$.
  \end{itemize}
  \item Each edge is associated with a propagator \emph{factor} $S(p,p')= \exp\left\{ -\int_p^{p'}\mu(x)\HS dx \right\}$, representing survival from age $p$ at the beginning of the edge, to age $p'$ at the end.
  \item The $j^\textrm{th}$ internal node is associated with birth \emph{factor} $\beta(p')$, where $p'$ is the age associated with the end of the edge below the node.
  \item All \emph{factors} are multiplied and integrated with respect to the measure $\displaystyle\int_{(\mathbb{R}^{+})^I} \dd \p_I \int_\Delta \dd \t_J$, where $\Delta$ represents the simplex region $0 \le t_1 \le t_2 \le \cdots \le t_J \le T$.
\end{itemize}


\subsection*{Fission Birth}

The construction of Feynman diagrams and corresponding Age-Time diagrams is required to represent terms arising in the expansion of $X_{m,n}(\q_m;\bar{q}_n;t)$ (see also Fig. \ref{FissFeyn}), each term begin an integral over a product of \emph{factors}. These are outlined below:

\begin{itemize}
  \item There are two types of initiating nodes ($I_1+I_2<m+n$):
  \begin{itemize}
  \item A number $I_1$ of circular initiating nodes,
\begin{tikzpicture}
\draw [black,thick] (0.1,0.1) circle [radius=0.1];
\draw [thick] (0.1,0.2) -- (0.1,0.28);
\end{tikzpicture}
, each associated with \emph{factor} $\alpha\omega(p_i)$, $i=1,...,I_1$, where $p_i$ represents the age of a founder singleton.
  \item A number $I_2$ of square initiating nodes,
\begin{tikzpicture}
\draw [black,thick] (0,0) rectangle (0.2,0.2);  
\draw [thick] (0.1,0.2) -- (0.1,0.28);
\end{tikzpicture}
, each associated with \emph{factor} $\bar{\alpha}\bar{\omega}(\bar{p}_i)$, $i=1,...,I_2$, where $\bar{p}_i$ represents the age of a founder pair of twins.
  \end{itemize}
  \item There are a number $J$ of internal nodes, two types of degrees $2$ (squares, 
\begin{tikzpicture}
\draw [black,fill=black] (0,0) rectangle (0.2,0.2);  
\draw [thick] (0.1,0.2) -- (0.1,0.28);
\draw [thick] (0.1,0.0) -- (0.1,-0.08);
\end{tikzpicture}  
, and circles,
\begin{tikzpicture}
\draw [black,fill=black] (0.1,0.1) circle [radius=0.1];
\draw [thick] (0.1,0.2) -- (0.1,0.28);
\draw [thick] (0.1,0.0) -- (0.1,-0.08);
\end{tikzpicture}
) and one of degree $3$ (hexagons, 
\begin{tikzpicture}
[scale=2]
\begin{scope}
    \foreach \x in {0,60,...,300} {
        \draw[fill=black] (0, 0) -- (\x:0.6 mm) -- (\x + 60:0.6 mm) -- cycle;}
\end{scope}
\draw [thick] (0.0,0.0) -- (0.0,-0.1);
\draw [thick] (-0.045,0.02) -- (-0.075,0.05);
\draw [thick] (0.045,0.02) -- (0.075,0.05);
\end{tikzpicture}
). Collectively, they are associated with times $0 \le t_1 \le t_2 \le\dots \le t_J \le T$.
  \item There are two types of terminating nodes, all associated with time $T$. There are $m$ circular nodes,
\begin{tikzpicture}
\draw [black,thick] (0.1,0.1) circle [radius=0.1];
\draw [thick] (0.1,0.0) -- (0.1,-0.08);
\end{tikzpicture}
associated with singletons, where the $\ell^\textrm{th}$ is associated with age $q_\ell$, $\ell=1,2,\dots,m$.  There are $n$ square nodes,
\begin{tikzpicture}
\draw [black, thick] (0,0) rectangle (0.2,0.2);  
\draw [thick] (0.1,0.0) -- (0.1,-0.08);
\end{tikzpicture}
associated with singletons, where the $\ell^\textrm{th}$ is associated with age $\bar{q}_\ell$, $\ell=1,2,\dots,n$. For any terminating node with corresponding age $q'$ we take the first node connected below that is that does \emph{not} involve any edge to the right of a node of degree $3$.
  \begin{itemize}
  \item If it is the $j^\textrm{th}$ internal node, we associate \emph{factor} $\delta(q' - (T-t_j))$.
  \item If it is an initiating node with associated age $p''$, we associate \emph{factor} $\delta(q' - (T+p''))$.
  \end{itemize}
  \item There are three classes of edges bridging nodes with corresponding initial and final times $t$ and $t'$ where:
  \begin{itemize}
  \item Edges of the form
   \begin{tikzpicture}
   \draw [white] (0, 0) -- (1,0);
   \draw (0, 0.12) -- (1,0.12);
   \draw (0, 0.06) -- (1,0.06);
   \end{tikzpicture}
   represent a pair of twins at time $t$ that survive through to time $t'$. 
  \item Edges of the form
   \begin{tikzpicture}
   \draw [white] (0, 0) -- (1,0);
   \draw [dashed] (0, 0.12) -- (1,0.12);
   \draw (0, 0.06) -- (1,0.06);
   \end{tikzpicture}
   represent a pair of twins at time $t$ that become a singleton by time $t'$ (one twin dies).
   \item Edges of the form
   \begin{tikzpicture}
   \draw [white] (0, 0) -- (1,0);
   \draw (0, 0.09) -- (1,0.09);
   \end{tikzpicture}
   represent a singleton at time $t$ that is a singleton at time $t'$ (the singleton survives).
  \end{itemize}
  \item The nodes that can be connected by these edges are restricted, depending upon whether they represent singeltons or twins.
  \begin{itemize}
  \item Edges of the form
  \begin{tikzpicture}
   \draw [white] (0, 0) -- (1,0);
   \draw (0, 0.12) -- (1,0.12);
   \draw (0, 0.06) -- (1,0.06);
   \end{tikzpicture}
   or
   \begin{tikzpicture}
   \draw [white] (0, 0) -- (1,0);
   \draw [dashed] (0, 0.12) -- (1,0.12);
   \draw (0, 0.06) -- (1,0.06);
   \end{tikzpicture}
   can start from any node representing the birth of twins: 
   \begin{tikzpicture}
   \draw [black,thick] (0,0) rectangle (0.2,0.2);  
   \draw [thick] (0.07,0.2) -- (0.07,0.28);
   \draw [thick] (0.13,0.2) -- (0.13,0.28);
   \end{tikzpicture}, 
   \begin{tikzpicture}
   \draw [black,fill=black] (0.1,0.1) circle [radius=0.1];
   \draw [thick] (0.07,0.2) -- (0.07,0.28);
   \draw [thick] (0.13,0.2) -- (0.13,0.28);
   \end{tikzpicture},
   \begin{tikzpicture}
   \draw [black,fill=black] (0,0) rectangle (0.2,0.2);  
   \draw [thick] (0.07,0.2) -- (0.07,0.28);
   \draw [thick] (0.13,0.2) -- (0.13,0.28);
   \end{tikzpicture}, or
   \begin{tikzpicture}
[scale=2]
\begin{scope}
    \foreach \x in {0,60,...,300} {
        \draw[fill=black] (0, 0) -- (\x:0.6 mm) -- (\x + 60:0.6 mm) -- cycle;}
\end{scope}
\draw [thick] (-0.035,0.00) -- (-0.085,0.04);
\draw [thick] (-0.015,0.01) -- (-0.065,0.06);
\end{tikzpicture}
   (note the edge is from the left side of a hexagonal node). 
   \item Edges of the form
   \begin{tikzpicture}
   \draw [white] (0, 0) -- (1,0);
   \draw (0, 0.12) -- (1,0.12);
   \end{tikzpicture}
   can start from any node representing singletons: 
   \begin{tikzpicture}
   \draw [black,thick] (0.1,0.1) circle [radius=0.1];
   \draw [thick] (0.1,0.2) -- (0.1,0.28);
   \end{tikzpicture}, 
or
   \begin{tikzpicture}
[scale=2]
\begin{scope}
    \foreach \x in {0,60,...,300} {
        \draw[fill=black] (0, 0) -- (\x:0.6 mm) -- (\x + 60:0.6 mm) -- cycle;}
\end{scope}
\draw [thick] (0.025,0.005) -- (0.075,0.05);
\end{tikzpicture}
   (note the edge is from the right side of a hexagonal node). 
   \item Edges of the form
   \begin{tikzpicture}
   \draw [white] (0, 0) -- (1,0);
   \draw (0, 0.12) -- (1,0.12);
   \end{tikzpicture}
   or
   \begin{tikzpicture}
   \draw [white] (0, 0) -- (1,0);
   \draw [dashed] (0, 0.12) -- (1,0.12);
   \draw (0, 0.06) -- (1,0.06);
   \end{tikzpicture}
   can end at any node representing singletons: 
   \begin{tikzpicture}
   \draw [black,thick] (0.1,0.1) circle [radius=0.1];
   \draw [thick] (0.1,0.0) -- (0.1,-0.07);
   \end{tikzpicture}, 
or
   \begin{tikzpicture}
   \draw [fill = black,thick] (0.1,0.1) circle [radius=0.1];
   \draw [thick] (0.1,0.0) -- (0.1,-0.07);
   \end{tikzpicture}.
 \item Edges of the form
  \begin{tikzpicture}
   \draw [white] (0, 0) -- (1,0);
   \draw (0, 0.12) -- (1,0.12);
   \draw (0, 0.06) -- (1,0.06);
   \end{tikzpicture}
   can end at any node that is terminal or representing the birth from a twin parent: 
   \begin{tikzpicture}
   \draw [black,thick] (0,0) rectangle (0.2,0.2);  
   \draw [thick] (0.07,0.0) -- (0.07,-0.07);
   \draw [thick] (0.13,0.0) -- (0.13,-0.07);
   \end{tikzpicture}, 
   \begin{tikzpicture}
   \draw [fill=black,thick] (0,0) rectangle (0.2,0.2);  
   \draw [thick] (0.07,0.0) -- (0.07,-0.07);
   \draw [thick] (0.13,0.20) -- (0.13,-0.07);
   \end{tikzpicture}, or
   \begin{tikzpicture}
[scale=2]
\begin{scope}
    \foreach \x in {0,60,...,300} {
        \draw[fill=black] (0, 0) -- (\x:0.6 mm) -- (\x + 60:0.6 mm) -- cycle;}
\end{scope}
   \draw [thick] (0.02,0.0) -- (0.02,-0.09);
   \draw [thick] (-0.02,0.0) -- (-0.02,-0.09);
   \end{tikzpicture}
    \end{itemize}
  \item The Age-Time diagram is constructed as follows. Each edge connecting a pair of nodes in the Feynman diagram is associated with initial and final age-time pairings $(p,t)$ and $(p',t')$, respectively. There is a corresponding line from $(p,t)$ to $(p',t')$ in the Age-Time diagram (Fig. \ref{FissFeyn}E). The pairings can be calculated by working from the initiating nodes, upwards, with the following observations.
  \begin{itemize}
  \item Edges connected to the $i^\textrm{th}$ initial singleton or twin node are given an initial age-time pairing of $(p_i,0)$ or $(\bar{p}_i,0)$, respectively.
  \item If the starting node of an edge is the $j^\textrm{th}$ internal node, which also has degree $2$, the initial age-time pairing is $(0,t_j)$.
  \item If the starting node of an edge is the $j^\textrm{th}$ internal node, which also has degree $3$, and the edge extends to the left from the node, the initial age-time pairing is $(0,t_j)$ (representing birth of a new individual).
  \item If the starting node of an edge is the $j^\textrm{th}$ internal node, which also has degree $3$, and the edge extends to the right from the node, the initial age-time pairing is $(p'',t_j)$, where $p''$ is the final age of the edge below the node (representing the age of the parent who gave birth to a new individual).
  \item The final age-time pairing for an edge where the final node is the $j^\textrm{th}$ internal node is $(p-t+t_j,t_j)$, where $(p,t)$ is the initial age-time pairing. 
  \item Edges connected to the $\ell^\textrm{th}$ terminating singleton or twin node are given a final age-time pairing of $(q_\ell,T)$ or $(\bar{q}_\ell,T)$, respectively.
  \end{itemize}
  \item Each edge is associated with a propagator \emph{factor} based upon the survival function $S(p,p')= \exp\left\{ -\int_p^{p'}\mu(x)\HS dx \right\}$, where $p$ and $p'$ are the initial and final ages associated with the edge, as follows:
  \begin{itemize}
  \item Edges of the form
   \begin{tikzpicture}
   \draw [white] (0, 0) -- (1,0);
   \draw (0, 0.12) -- (1,0.12);
   \draw (0, 0.06) -- (1,0.06);
   \end{tikzpicture}
   have a propagator $S(p,p')^2$ representing survival of a singleton. 
  \item Edges of the form
   \begin{tikzpicture}
   \draw [white] (0, 0) -- (1,0);
   \draw [dashed] (0, 0.12) -- (1,0.12);
   \draw (0, 0.06) -- (1,0.06);
   \end{tikzpicture}
   have a propagator $2S(p,p')(1-S(p,p'))$, representing survival of one of a pair of twins.
   \item Edges of the form
   \begin{tikzpicture}
   \draw [white] (0, 0) -- (1,0);
   \draw (0, 0.09) -- (1,0.09);
   \end{tikzpicture}
   have a propagator $S(p,p')$, representing survival of a pair of twins.
  \end{itemize}
  \item We have the following birth factors:
  \begin{itemize}
  \item If the $j^\textrm{th}$ internal node is of the form
  \begin{tikzpicture}
\draw [black,fill=black] (0.1,0.1) circle [radius=0.1];
\draw [thick] (0.1,0.2) -- (0.1,0.28);
\draw [thick] (0.1,0.0) -- (0.1,-0.08);
\end{tikzpicture}  
it is associated with birth \emph{factor} $\beta(p')$, where $p'$ is the age associated with the end of the edge below the node.
\item  If the $j^\textrm{th}$ internal node is of the form
\begin{tikzpicture}
\draw [black,fill=black] (0,0) rectangle (0.2,0.2);  
\draw [thick] (0.1,0.2) -- (0.1,0.28);
\draw [thick] (0.1,0.0) -- (0.1,-0.08);
\end{tikzpicture}
or
\begin{tikzpicture}
[scale=2]
\begin{scope}
    \foreach \x in {0,60,...,300} {
        \draw[fill=black] (0, 0) -- (\x:0.6 mm) -- (\x + 60:0.6 mm) -- cycle;}
\end{scope}
\draw [thick] (0.0,0.0) -- (0.0,-0.1);
\draw [thick] (-0.045,0.02) -- (-0.075,0.05);
\draw [thick] (0.045,0.02) -- (0.075,0.05);
\end{tikzpicture}
it is associated with birth \emph{factor} $2\beta(p')$, where $p'$ is the age associated with the end of the edge below the node.
\end{itemize}
\item All \emph{factors} are multiplied and integrated with respect to the measure $\displaystyle\int_{(\mathbb{R}^{+})^{I_1+I_2}} \dd \p_{I_1} \HS \dd \p_{I_2} \int_\Delta \dd \t_J$, where $\Delta$ represents the simplex region $0 \le t_1 \le t_2 \le \cdots \le t_J \le T$.
\end{itemize}


\setcounter{section}{6}
\section*{Appendix F: Exact Approach to Correlation Functions}
\addcontentsline{toc}{section}{Appendix F: Exact Approach to Correlation Functions}
\setcounter{equation}{0}
\label{AppExact}

The path integral in Eq. \ref{MOMPI} can be written (using integration by parts) as

\begin{eqnarray}
X_m(\q_m;T) =\left.\frac{\delta^mZ(K)}{\prod_{i=1}^m\delta K(q_i;T)}\right|_{K\equiv0},
\label{AFFD}
\end{eqnarray}
where
\begin{eqnarray}
Z(K)=\int \mathcal{D}u\mathcal{D}v& \exp\left\{i\int \dd q \HS\dd t\HS v\left[\left(\frac{\partial u}{\partial q}+\frac{\partial u}{\partial t}-\mu u\right)
+\beta (u+1) u^0 +K\right]\right\}\cdot\nonumber\\
& \hspace{5mm}\exp\left\{ - i\int \dd q \hspace{1mm}u_T v_T + \alpha\int \dd q \HS \omega u_0\right\}.
\end{eqnarray}

Now functional integration over the $v$ variable results in the constraint $\frac{\partial u}{\partial q}+\frac{\partial u}{\partial t}-\mu v +\beta (u+1) u^0 +K=0$ and integration over $v_T$ results in the boundary condition $u_T=0$. Now this can be solved with the method of characteristics to give an expression of the form $Z(K) = \exp\left\{\alpha\int \dd q \HS \omega u_0\right\}$ where $u_0(q)=u(q,0)$ is found from,

\begin{eqnarray}
u(q,t) & = & \int_0^{T-t}\dd s \HS\left[K(q+T-t,s) + \beta(q+T-t-s)B(s)\right]\cdot\nonumber\\
&& \exp\left\{-\int_s^{T-t}\dd s' \HS\left[\mu(q+T-t-s')-\beta(q+T-t-s')B(s')\right]\right\},
\end{eqnarray}
and the boundary term $B(t)=u(0,T-t)=u^0(T-t)$ satisfies the integral equation obtained when substituting $q \rightarrow 0$, $t\rightarrow T-t$. The fact that $u(q,t)$ and $B(t)$ have functional dependence upon $K$ makes the functional derivative of Eq. \ref{AFFD} difficult to implement, and finding a useful way to analyze $X_m(\q_m;t)$ or $f_m(\q_m;t)$ without perturbation remains an open problem.

\bibliographystyle{iopart-num} 
\bibliography{refs_DoiPeliti}

\end{document}